\documentclass[aip,graphicx, reprint]{revtex4-2}

\usepackage[dvipsnames]{xcolor}
\usepackage{bm}
\usepackage{graphicx}
\usepackage{upgreek}

\usepackage{natbib}
\usepackage{braket}
\usepackage{physics}
\usepackage[%
breaklinks,%
pdftex,%
hyperfootnotes=true,%
pdfpagelabels,%
bookmarks,%
pageanchor,%
]{hyperref}

\hypersetup{%
	colorlinks=true, linktocpage=true, pdfstartpage=1, pdfstartview=FitH, pdfborder={0 0 0},%
	breaklinks=true, pdfpagemode=UseNone, pageanchor=true, pdfpagemode=UseOutlines,%
	plainpages=false, bookmarksnumbered, bookmarksopen=true, bookmarksopenlevel=1,%
	hypertexnames=true, pdfhighlight=/O,
	urlcolor=Red!50!Sepia, linkcolor=NavyBlue, citecolor=RoyalBlue, 
}

\newcommand{\red}[1]{\textcolor{black}{#1}}

\newcommand{\ie}{i.e., }

\def\>{\right\rangle}
\def\<{\left\langle}
\def\be{\begin{equation}}
\def\ee{\end{equation}}
\def\ba{\begin{array}{lll}}
\def\ea{\end{array}}

\def\beq{\begin{eqnarray}}
\def\eeq{\end{eqnarray}}

\def\d{{\rm d}}

\begin{document}
\title{Dissipation-induced \red{collective advantage} of a quantum thermal machine}
\author{Matteo Carrega}
\email{matteo.carrega@spin.cnr.it}
\affiliation{CNR-SPIN,  Via  Dodecaneso  33,  16146  Genova, Italy}
\author{Luca Razzoli}
    \affiliation{Center for Nonlinear and Complex Systems, Dipartimento di Scienza e Alta Tecnologia, Universit\`a degli Studi dell'Insubria, via Valleggio 11, 22100 Como, Italy} 
    \affiliation{Istituto Nazionale di Fisica Nucleare, Sezione di Milano, via Celoria 16, 20133 Milano, Italy}
\author{Paolo Andrea Erdman}
\affiliation{Freie Universit{\" a}t Berlin, Department of Mathematics and Computer Science, Arnimallee 6, 14195 Berlin, Germany}
\author{Fabio Cavaliere}
    \affiliation{Dipartimento di Fisica, Universit\`a di Genova, Via Dodecaneso 33, 16146 Genova, Italy} 
    \affiliation{CNR-SPIN,  Via  Dodecaneso  33,  16146  Genova, Italy}
\author{Giuliano Benenti}
    \affiliation{Center for Nonlinear and Complex Systems, Dipartimento di Scienza e Alta Tecnologia, Universit\`a degli Studi dell'Insubria, via Valleggio 11, 22100 Como, Italy} 
    \affiliation{Istituto Nazionale di Fisica Nucleare, Sezione di Milano, via Celoria 16, 20133 Milano, Italy}
    \affiliation{NEST, Istituto Nanoscienze-CNR, P.zza San Silvestro 12, I-56127 Pisa, Italy}
    \author{Maura Sassetti}
    \affiliation{Dipartimento di Fisica, Universit\`a di Genova, Via Dodecaneso 33, 16146 Genova, Italy} 
    \affiliation{CNR-SPIN,  Via  Dodecaneso  33,  16146  Genova, Italy}

\begin{abstract}
\red{Do quantum correlations lead to better performance with respect to several different systems working independently?}
For quantum thermal machines, the question is whether a working medium (WM) made of $N$ constituents exhibits better performance than $N$ independent engines working in parallel. Here, by inspecting a microscopic model with the WM composed by two non-interacting quantum harmonic oscillators, we show that the presence of a common environment can mediate non-trivial correlations in the WM leading to \red{better quantum heat engine performance---maximum power and efficiency---
with respect to an independent configuration.}
\red{Furthermore, this advantage} is striking for strong dissipation, a regime in which two independent engines cannot deliver any useful power. Our results show that dissipation can be exploited as a useful resource for quantum thermal engines, and are corroborated by \red{optimization techniques} here extended to non-Markovian quantum heat engines. 
\end{abstract}
\maketitle
\section{Introduction}
\red{On the road to} the development of quantum technologies~\cite{acin18}, a fundamental question is \red{whether quantum correlations between the constituents of a system can improve performances}~\cite{qcbook}. (for instance, the qubits in a quantum computer). In the context of quantum thermal machines~\red{\cite{kosloff13,kurizki15,sothmann15,anders15,goold16,benenti17,jurgen,paternostro21,pekola21,arracheaR, polettirmp, cangemi_bhadra, rosa0, rosa1, sing, lili, cangemi_prr, stock}}, the question can be posed as follows:
can a working medium (WM) made of $N$ constituents show better performances than $N$ independent engines working in parallel?
Furthermore, can unavoidable dissipation be exploited to improve machine performance or does it only play a detrimental role?
Even though previous investigations remarkably found parameter regions where \red{a positive answer~\cite{Campisi2016,Jaramillo2016,Vroylandt2018,Hardal2018,Niedenzu2018,GelbwaserKlimovsky2019,Latune2020,gonzalo_2019, victor, Watanabe2020,Souza2022,tajima, Kamimura2022,Kamimura2023}, even related to damping induced phenomena~\cite{paz09, galve10, correa12, duarte, henriet, correa13, valido15, henkel21, feyisa2023, marino, rolandi, hardal} can be given, a complete picture has not yet been achieved}. 
\begin{figure}[ht]
    	\centering
 \includegraphics[width=0.9\linewidth]{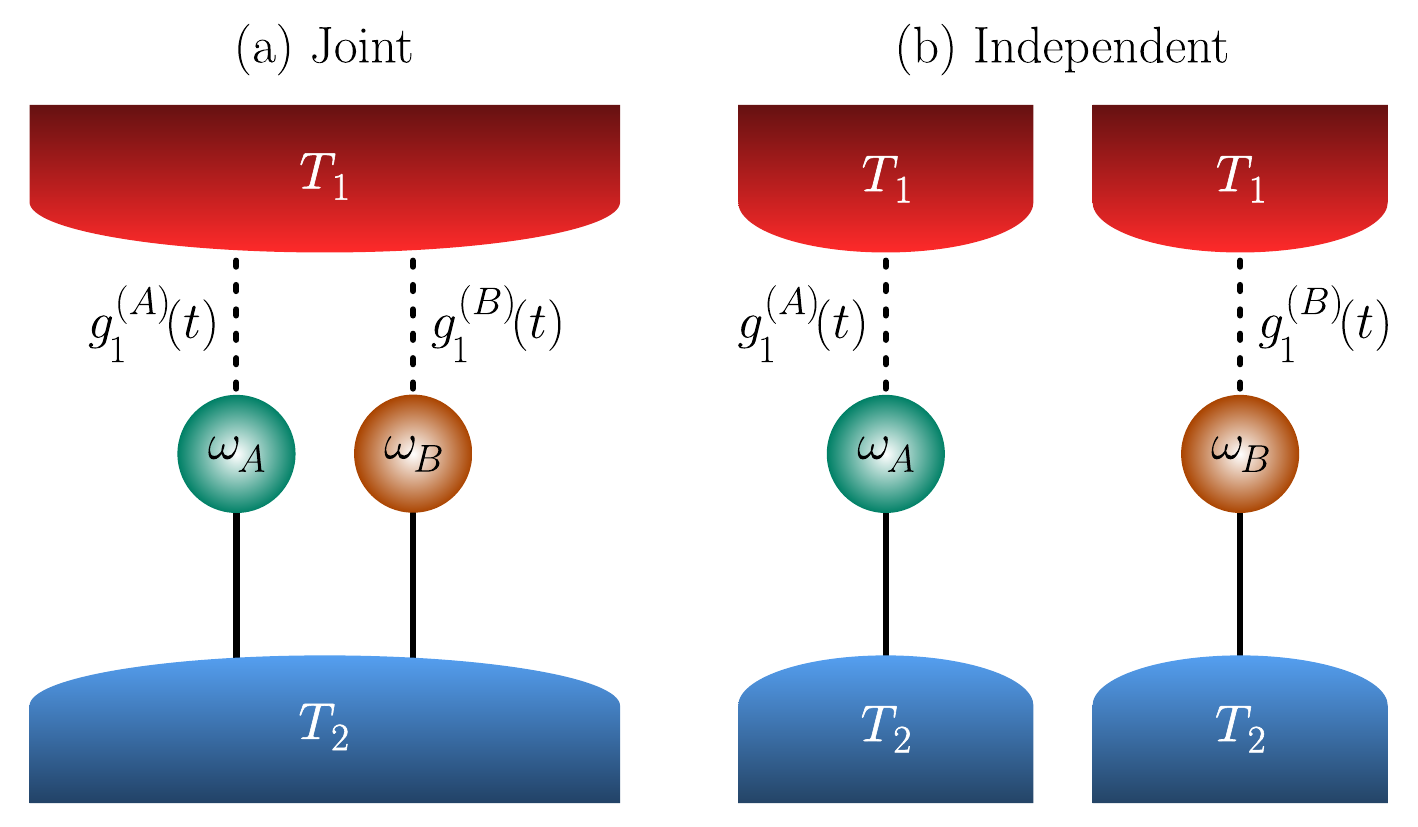}
\caption{Sketch of the dynamical quantum heat engines under study. The external driving of the machine occurs through \emph{generic} periodic modulations of the couplings with one reservoir. (a) Two quantum harmonic oscillators, with frequencies $\omega_A$ and $\omega_B$, are in contact with two common thermal reservoirs at temperatures $T_\nu$, with $\nu =1,2$. The WM exchanges heat
currents $J_\nu$ with the reservoirs and total power $P$ generated by an external drive  that periodically modulates the weak coupling with the $\nu=1$ reservoir (dashed lines).
 The $\nu=2$ WM-bath coupling is static and much stronger (solid lines). (b) Two uncoupled oscillators in the same configuration as in Panel (a) but now in contact with independent thermal reservoirs. Here, no correlations are mediated by the baths, and the quantum thermal machine consists of independent two-terminal devices working in parallel.
\label{fig:1}}
\end{figure}
In this work, we show that a quantum heat engine where the working medium is composed by two non-interacting quantum harmonic oscillators, connected to common baths -- see Fig.~\ref{fig:1}(a) -- via periodically modulated couplings~\cite{carrega_prxquantum, cavaliere_prr, cavaliere_iscience}, exhibits \red{an improvement in performance} due to bath-mediated correlations~\cite{paz09, correa12} with respect to the case of two independent single-oscillator engines working in parallel -- see Fig.~\ref{fig:1}(b). \red{This effect} is striking in the case of strong dissipation, when one would naively expect overdamped dynamics and poor performance of the engine. While this is the case for independent machines, for which the engine operating regime disappears, the presence of common baths sustains instead efficiency and power of the engine. We explain this surprising result in terms of the appearance, at strong damping, of a frequency-- and phase--locked mode~\cite{Giorgi_2012,Manzano_2013,Du_2017,Geng_2018}, in which the oscillators have a \red{common} frequency and  oscillate in phase opposition.
This \red{normal} mode turns out to be only weakly damped, with a damping time increasing with the dissipation strength. 
Our results are first illustrated in the case of monochromatic drives, and then corroborated optimizing over arbitrary periodic drivings by means of \red{ a gradient optimization methods (a technique at the heart of many machine learning problems~\cite{ashida2021, sgroi2021, abiuso_prl, martin_entropy, erdman2023_prr, erdman2023_pnas, erdman2022, khait2022})}, without any a priori assumption on the shape or speed of the drivings. Furthermore, we characterize the heat engine performance in terms of both efficiency and extracted power. These quantities cannot be simultaneously optimized: indeed, high power engines typically exhibit a low efficiency, and vice-versa. 
We thus employ the concept of the Pareto front~\cite{Pareto}, recently employed in the context of 
quantum thermodynamics~\cite{patel2017, solon2018, gonzalez2020, ashida2021, erdman2023_prr, erdman2023_pnas}, to find optimal tradeoffs between the power and the efficiency of the engine. 
Our calculations of the Pareto front \red{extend the optimization approach} to non-Markovian quantum thermodynamics, beyond the standard Lindblad approximation.
Finally, we have investigated whether there is a relationship between \red{collective} advantage and the establishment of non-classical correlations between the two quantum harmonic oscillators (QHOs), focusing
on the logarithmic negativity~\red{\cite{correa12, Vidal2002, plenio_prl, simon2000, serafini2004, paz_roncaglia, rebecca}} as a measure of entanglement. While we could not find a direct connection between entanglement and \red{collective} advantage,
we obtained as an interesting byproduct a protocol to measure entanglement via a small number of measurements of thermodynamic quantities, more precisely of the output work at specific operating conditions, instead of a full quantum tomography to reconstruct the density matrix of the system.
\section{General setting and thermodynamic observables}
\subsection{\red{Model}}
We consider a quantum thermal machine, where the WM is in contact with two thermal reservoirs $\nu=1,2$ respectively at temperatures $T_1$ and $T_2$. The WM is composed of two uncoupled (no direct coupling) QHOs, labelled $l=A,B$, as sketched in Fig.~\ref{fig:1}(a). This configuration with common environments, dubbed {\it joint}, will be compared with the one sketched in Fig.~\ref{fig:1}(b), where two independent QHOs work in parallel with separate baths.
The WM-bath couplings with the $\nu=1$ reservoir are assumed weak and governed by a time--dependent periodic modulation~\cite{carrega_prxquantum, cavaliere_prr, cavaliere_iscience} $g_1^{(l)}(t)=g_1^{(l)}(t+{\cal T})$ with period ${\cal T}=2\pi/\Omega$. On the other hand, the couplings with the $\nu=2$ reservoir are  static, $g_2^{(l)}=1$. Coupling modulation can be suitably engineered to perform thermodynamic tasks~\cite{carrega_prxquantum,cavaliere_prr, cavaliere_iscience}, and here we shall focus on the heat engine working mode.

The total Hamiltonian is (we set $\hbar = k_B = 1$)~\cite{note_superscript}
\begin{equation}
H^{(t)} = \sum_{l=A,B} H_{l} + \sum_{\nu=1,2} \big[ H_\nu + H_{{\rm int}, \nu}^{(t)}\big]~,
\label{eq:totH}
\end{equation}
where the Hamiltonian of the $l$-th QHO reads $H_l= \frac{p_l^2}{2 m} + \frac{1}{2}m \omega_l^2 x_l^2$, with the two QHOs having different characteristic frequencies $\omega_A$ and $\omega_B$.
The reservoirs are modelled in the Caldeira-Leggett framework~\cite{caldeira1983, weiss, weiss2, aurel18}  as a collection of independent harmonic oscillators
\red{\begin{equation}
H_{{\nu}}=  \sum_{k=1}^{+\infty} \left(\frac{P^2_{k,\nu}}{2 m_{k,\nu}} + \frac{1}{2}m_{k,\nu} \omega^2_{k,\nu }X^2_{k,\nu}\right)~.
\label{eq:Ham_res}
\end{equation} }
A bilinear coupling in the WM and bath position operators, weighted by the driving controls $g_\nu^{(l)}(t)$, describes the WM-reservoir interactions
\red{\beq
&& H_{{\rm int},\nu}^{(t)} = \sum_{l=A,B} \sum_{k=1}^{+\infty} \bigg[
- g_\nu^{(l)}(t) c_{k,\nu}^{(l)} x_l X_{k,\nu}
\nonumber \\
&&+\frac{(g_\nu^{(l)}(t) c_{k,\nu}^{(l)})^2}{2 m_{k,\nu}\omega_{k,\nu}^2} x_{l}^2 
+ \frac{g_\nu^{(l)}(t) g_\nu^{(\bar{l})}(t)  c_{k,\nu}^{(l)} c_{k,\nu}^{(\bar{l})} }{2 m_{k,\nu}\omega_{k,\nu}^2} x_{l}x_{\bar{l}} \bigg]~,
\label{eq:H_int_nu}
\eeq
}\red{where we introduced the convention according to which if $l=A$ then $\bar{l} = B$, and \textit{vice versa}.
The factors $c_{k,\nu}^{(l)}$ represent the \red{coupling} between the $l$-th QHO and the $k$-th mode of the $\nu$-th reservoir. In the following we assume that the couplings with the bath $\nu=2$ are much stronger than those with the bath $\nu=1$. Without loss of generality, we also choose for the bath $\nu$ equal couplings $
c_{k,\nu}^{(A)}=c_{k,\nu}^{(B)}\equiv c_{k,\nu}\,$.} \red{Looking ath the coupling with the static bath $\nu=2$, this choice of equal couplings leads to a mirror symmetry $A\leftrightarrow B$. In the resonant case $\omega_A=\omega_B$ this symmetry explains the existence of a dissipation-free subspace, with the normal mode corresponding to relative coordinate $x_A-x_B$ completely undamped~\cite{paz09, correa12, duarte}. However, this implies that in the resonant case the system cannot reach a periodic steady state, and thus we are not going to deal with this case in the rest of this work.} 
\red{The interaction in Eq.~\eqref{eq:H_int_nu} includes counter-term contributions that serve two purposes: to avoid renormalizations of the characteristic frequencies of the QHOs $\omega_{A,B}$ and to cancel the direct coupling among them, that would naturally arise in the Caldeira--Leggett model (see \red{supporting information} SI).
The properties of the bath $\nu$ are governed by the so-called spectral density~\cite{weiss}
\be
\label{eq:spectral}
{\cal J}_\nu(\omega)\equiv\frac{\pi}{2}\sum_{k=1}^{+\infty} \frac{c^{2}_{k,\nu}}{m_{k,\nu}\omega_{k,\nu}}\delta(\omega-\omega_{k,\nu})~.
\ee
}
\red{
Finally, we assume that at initial time $t_0{\to -\infty}$, the reservoirs are in their thermal equilibrium at temperatures $T_\nu$, with the total density matrix written in a factorized form 
$\rho(t_0) = \rho_{A}(t_0)\otimes\rho_{B}(t_0)\otimes \rho_{1}(t_0)\otimes \rho_{2}(t_0)$, where $\rho_{l}(t_0)$ is the initial density matrix of each QHO ($l=A,B$), and $\rho_\nu (t_0) = \exp(-H_\nu / T_\nu)/\Tr[\exp(-H_\nu / T_\nu)]$
is the thermal density matrix of each reservoir ($\nu=1,2$).
}
\subsection{\red{Thermodynamic quantities}} 
\red{Hereafter, we work in the Heisenberg picture, and we} focus on averaged thermodynamic quantities such as power and heat currents, which determine the working regime and the performance of a quantum thermal machine. Except in the resonant case $\omega_A\equiv\omega_B$~\cite{paz09, correa12}, due to dissipation the WM reaches a periodic steady state regardless of the initial conditions.
We then concentrate on quantities averaged over the period ${\cal T}$, in the off resonant case $\omega_B < \omega_A$. The average power is defined as
\be
\label{averagepower}
P\equiv \int_0^{{\cal T}}\frac{\d t}{{\cal T}} \langle P(t)\rangle =\int_0^{{\cal T}}\frac{\d t}{{\cal T}} {\rm Tr}\left[\frac{\partial H_{{\rm int},1}^{(t)}(t)}{\partial t}\rho(t_0)\right]~,
\ee
where we have introduced both the temporal and quantum averages (the latter denoted by $\langle\ldots\rangle$), $t_0\to -\infty$ is the initial time and $\rho(t_0)$ the initial density matrix of the system (see App.~\ref{app_b}). Notice that a working heat engine is obtained when $P<0$.
Similarly, the average heat current associated to the $\nu$-th reservoir reads
\be
\label{averagecurrent}
J_\nu \equiv \int_0^{{\cal T}}\frac{\d t}{{\cal T}}\langle J_\nu(t)\rangle=-\int_0^{{\cal T}}\frac{\d t}{{\cal T}}{\rm Tr}\left[\dot{H}_\nu(t)\rho(t_0)\right],
\ee
with $J_\nu>0$ when energy flows into the WM. The average power and heat currents are expressed (see App.~\ref{app_b}) in terms of the QHOs and bath position operators $x_l(t)$ and $X_{k,\nu}(t)$, respectively.
The exact solution for $X_{k,\nu}(t)$ can be found by inspecting the set of coupled equations of motion, see \red{also  SI}. The behaviour of thermodynamic quantities is eventually determined by the dynamics of $x_A(t)$ and $x_B(t)$.

As discussed in Ref.~\cite{cavaliere_prr}, the above quantities satisfy the energy balance relation $P+\sum_\nu J_\nu=0$, in compliance with the first law of thermodynamics. Another relevant quantity of interest is the so-called entropy production rate 
$\sigma\equiv -\sum_\nu\frac{J_\nu}{T_\nu}$. In accordance with the second law of thermodynamics, it is always~\cite{esposito2010, paternostro21} $\sigma \ge 0$. This represents another key figure of merit for thermal machines: for instance for a good heat engine one should look for the best power output while minimizing at the same time the entropy production rate.

\section{Results}
\subsection{Bath-induced dynamics and response functions}
Under the assumption that the WM is weakly coupled to the modulated $\nu=1$ reservoir, a perturbative approach in $H_{{\rm int},1}^{(t)}$ is considered. The final expressions for the average power and heat current read (see SI) 
\be
\label{eq:p_final}
P \! = \! - \!\!\!\!\sum_{n=-\infty}^{+\infty} \!\!\!  n\Omega \!\! \int_{-\infty}^{+\infty} \!\!\! \frac{\d \omega}{2\pi m} {\cal J}_1(\omega+n\Omega) N(\omega,n\Omega) {\bf g}^\dagger _n\cdot {\bm \chi}_2''(\omega)\cdot {\bf g}_n , 
\ee
and 
\begin{align}
\label{eq:j1_final}
\! J_1 \! = \!  \sum_{n=-\infty}^{+\infty}  \int_{-\infty}^{+\infty} & \frac{\d\omega}{2\pi m}  (\omega + n\Omega) {\cal J}_1(\omega+n\Omega) N(\omega,n\Omega)  \nonumber\\
&\times {\bf g}^\dagger _n\cdot{\bm \chi}''_2 (\omega)\cdot{\bf g}_n  ,
\end{align}
and $J_2= - (P+J_1)$. In the above expressions \red{enter ${\cal J}_1(\omega)$, the spectral density of the $\nu=1$ bath }, governing memory effects, and the function 
\begin{equation}
\label{eq:n_function}
N(\omega, \Omega) = \coth\left(\frac{\omega + \Omega}{2T_1}\right)-\coth\left(\frac{\omega}{2T_2}\right)~.
\end{equation}
\red{It is worth noticing that the above quantities are written as quadratic forms where we have introduced the $2n$ components vector ${\bf g}_n=(g_n^{(A)}, g_n^{(B)})^{t}$ (and ${\bf g}_n^\dagger$ its adjoint) with $g_n^{(l)}$ the $n$--th Fourier coefficient of $g_1^{(l)}(t)$.}
\red{We assume that these coefficients are related to two independent drive sources, hence they satisfy two independent constraints
\be
\label{gnl}
	\sum_n |g_n^{(l)}|^2 = {g^{(l)}}^2,\, 
	\text{for } l=A,B,
\ee
where $g^{(l)}$ are two fixed normalization constants.}
Both $P$ and $J_\nu$ depend on the imaginary part, ${\bm \chi}_2''(\omega)$, of the retarded response matrix ${\bm \chi}_2(\omega)= {\bm \chi}_2'(\omega) + i {\bm \chi}_2''(\omega)$. The elements of the two-by-two matrix are indeed the Fourier transform of ($l,l'= A,B$)
\be
\label{eq:responsefunction}
\chi^{(l,l')}_2(t)\equiv i m \theta(t)\langle[x^{(0)}_l(t), x_{l'}^{(0)}(0)]\rangle~,
\ee
where $\theta(t)$ is the step function and $x_l^{(0)}(t)$ evolve under the unperturbed Hamiltonian and their response functions are linked to the static $\nu=2$ bath only (see App.~\ref{app_c}). Their imaginary parts read 
\begin{align}\label{eq:chischis}
\!\!\!&\!\!\!{\chi_2^{(l l)}}''(\omega)=\frac{\omega(\omega^2-\omega_{{\bar l}}^2)^2\gamma_2'(\omega)}{| D(\omega)|^2},\nonumber \\
&{\chi_2^{(l {\bar l})}}''(\omega) =\frac{\omega(\omega^2-\omega_l^2)(\omega^2-\omega_{{\bar l}}^2)\gamma_2'(\omega)}{|D(\omega)|^2}~,
\end{align}
\red{
\be
\label{eq:denominator2}
D(\omega)\!=\!(\omega^2\!- \omega_A^2)(\omega^2\!- \omega_B^2) + i\omega(2 \omega^2\! - \omega_A^2 - \omega_B^2)\gamma_2(\omega),
\ee
}
and where $\gamma'_2(\omega)$ is the real part of the damping kernel $\gamma_2(\omega)$ of the bath $\nu=2$ (see App.~\ref{app_c}). 
In the following discussion, we will assume a Ohmic spectral density for the latter bath, which implies $\gamma_2(\omega)=\gamma_2$, a constant real number \cite{weiss}. 
\red{Looking at the structure of the  symmetric matrix ${\bm \chi}_2''(\omega)$, one can notice that at any frequency $\omega$ there is a null eigenvalue and a finite one given by 
\be
\lambda=\frac{\gamma_2\omega}{|D(\omega)|^2}\big[2\omega^4-2(\omega_A^2+\omega_B^2)\omega^2 +\omega_A^4+\omega_B^4\big]~.
\ee
}
\red{It turns out that its associated frequency-dependent eigenvector, when evaluated at $\omega_{A/B}$ corresponds to the {\it localized} vector $(1,0)^t \ / \ (0,1)^t$. Furthermore, it is possible to find a completely {\it delocalized} and anti-symmetric eigenvector $(-1,1)^t$. This is achieved when the frequency is equal to
\be
\label{eq:hybrid}		
\bar{\omega}=\sqrt{\frac{\omega_A^2+\omega_B^2}{2}},
\ee
whose value will play an important role (see below).}
\red{It is instructive to evaluate the imaginary part of the response function at these three points:
\beq
{\bm \chi}_2''(\omega_l) =\frac{1}{\gamma_2\omega_l}\left(\begin{array}{cc}
\delta_{l,A} &0\\
0&\delta_{l,B}\end{array}\right)
\eeq
and
\beq
{\bm \chi}_2''(\bar{\omega})=\frac{\gamma_2\bar{\omega}}{\Delta^4}\left(\begin{array}{cc}
1 &-1 \\
-1 &1\end{array}\right)
\eeq
where we introduced $\Delta=\sqrt{(\omega_A^2-\omega_B^2)/2}$ for notational convenience. From the above expression a completely different behaviour as a function of $\gamma_2$ emerges. Indeed, for $\gamma_2 \ll \omega_l$ the localized modes are predominant, while they subside for $\gamma_2 \gg \omega_l$, when the delocalized mode becomes the leading one, as can be seen by the prefactors of the above matrices. In Fig.~\ref{fig:eigenvalue} we plot the finite eigenvalue $\lambda$ in a density plot as a function of frequency $\omega$ and damping strength $\gamma_2$. From this picture the above structure, and its evolution with increasing damping strength,  becomes clear.} 

 \red{
  \begin{figure} 
 \centering
\includegraphics[width=0.9\linewidth]{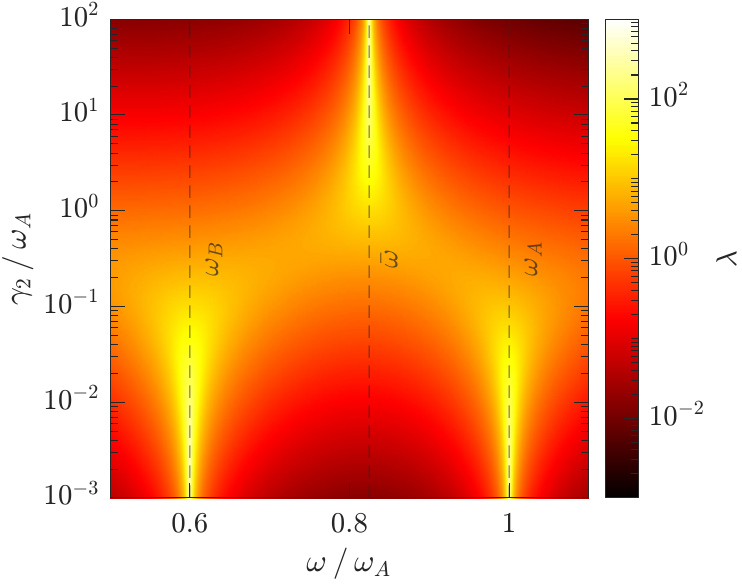}
\caption{Density plot of the eigenvalue $\lambda$ as a function of frequency $\omega$ and damping strength $\gamma_2$ (in units of $\omega_A$). We have fixed $\omega_A=1$ and $\omega_B=0.6$. Notice that with this choice $\bar{\omega}$ of Eq.~\eqref{eq:hybrid} corresponds to $0.82$. 
\label{fig:eigenvalue} }
\end{figure}
}
\red{This shows that the system response crucially depends on the damping strength $\gamma_2$. More precisely, this can be seen by inspecting 
the zeros of $D(\omega)$ that govern intrinsic excitations of the normal modes (see also App.~\ref{app_c}).}
 Explicitly,  at very weak damping (\ie when $\gamma_2\ll\omega_l$)  ${\bm \chi}_2''(\omega)$ \red{resembles} the one of two independent QHOs \red{with differences of order $O(\gamma_2/\omega_l$)}
Conversely, in the opposite strong damping regime (\ie when $\gamma_2\gg \Delta^4/\bar{\omega}^3$) the key result is that the WM becomes effectively {\it frequency locked} to a unique characteristic frequency.
In this regime the two QHOs oscillate, at finite time, with a \red{common} frequency $\bar{\omega}$. Moreover, they are also {\it phase locked} in anti--phase (see App.~\ref{app_c}). This important behaviour is tightly related to {\em bath-mediated correlations}: indeed, $\gamma_2$ plays a twofold role. On the one hand it is responsible for dissipation but, on the other hand, it also mediates an effective coupling between the two QHOs, \red{establishing} non-trivial correlations between them even without any direct, a--priori coupling.

\subsection{Quantum thermal machine performance}
We now present the effect of bath-mediated correlations on the performance of a dynamical heat engine. To ensure a working heat engine~\cite{cavaliere_prr, cavaliere_iscience} we choose for the bath $ \nu=1$ a structured non-Markovian environment~\red{\cite{thorwart, paladino, nazir14, groeblacher, illuminatiprl, note_ohmicbath}} with a Lorentzian spectral function
\be 
{\cal J}_1(\omega)= \frac{d_1 m\gamma_1 \omega}{(\omega^2 - {\omega}_1^2)^2 + \gamma_1^2\omega^2}~\label{eq:lorspd},
\ee
with a peak centered at $\omega\sim {\omega}_1$, an amplitude governed by  $d_1$, and a width determined by $\gamma_1$. Notice that for sufficiently small $\gamma_1$ this spectral density acts as a sharp filter, centered around $\pm \omega_1$. Such structured environment represents a common example of non-Markovian bath~\cite{groeblacher, cavaliere_prr} and 
can be realized with state-of-the-art superconducting circuits~\cite{cottet, calzona, peropadre, rodrigues}.
\begin{figure*} 
 \centering
\includegraphics[width=0.9\linewidth]{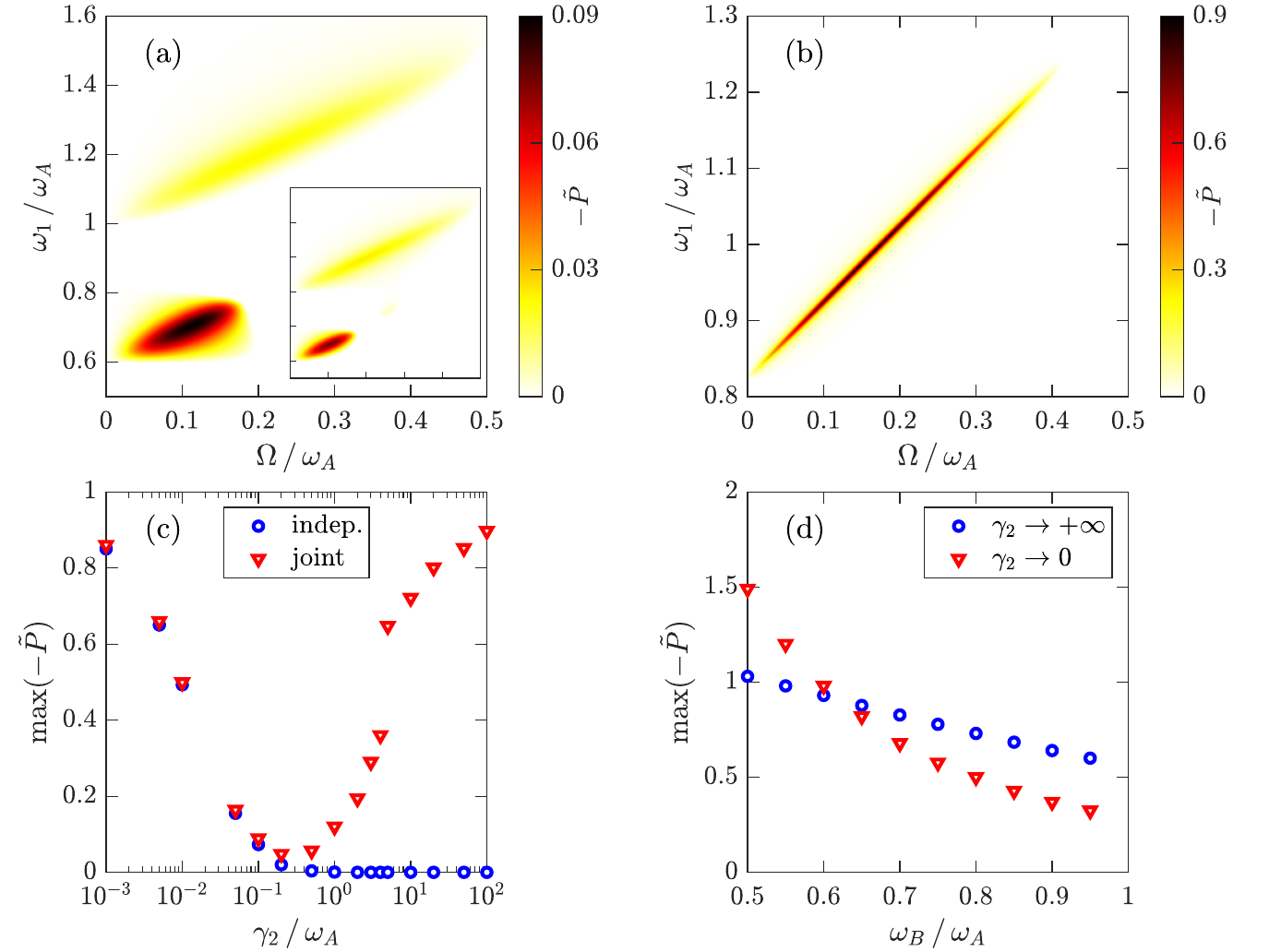}
        \caption{Output power of Eq.~\eqref{eq:p_singleharmonics} for the dynamical heat engine with a monochromatic drive. (a) Density plot of the (dimensionless) output power $-\tilde{P} = - P\omega_A^2/d_1$ for the joint configuration, obtained by optimizing with respect to the phase displacement $\phi$, as a function of $\Omega/\omega_A$ and $\omega_1/\omega_A$ with $\omega_B=0.6 \omega_A$, $\phi=0$, and damping strength $\gamma_2=0.1\omega_A$. The inset shows the comparison with the independent configuration of Fig.~\ref{fig:1}(b), on the same scale. (b) Same as in panel (a) but for strong dissipation  with $\gamma_2=100\omega_A$ and $\phi=\pi$ (see text). (c) Maximum output power for the representative value $\omega_B=0.6\omega_A$ as a function of damping strength $\gamma_2$, both in the joint and in the independent case. For the joint case, power is maximized over $\omega_1$, $\Omega$, and $\phi$. (d) Maximum output power for the joint configuration in the very weak ($\gamma_2\to 0$) and ultra--strong ($\gamma_2\to \infty$) 
regime, as a function of $\omega_B/\omega_A$. Here, the same optimization as in Panel (c) has been performed. Other parameters are: $ T_1=0.6\omega_A$, $T_2=0.4\omega_A$, and $\gamma_1=0.01\omega_A$. 
        \label{fig:density}}
\end{figure*}

To begin, we investigate the output power produced by the heat engine, in the \red{simple }case of a monochromatic drive: $g_1^{(A)}(t)=\cos(\Omega t)$ and $g_1^{(B)}(t)=\cos(\Omega t+\phi)$, with $\Omega$ the external frequency and $\phi$ the relative phase \red{of two independent} drives.
Although the choice of a single harmonic might seem a simplifying assumption, below we show that in most cases this represents the optimal one.
In this case,\red{ in order to enforce the two constraints in Eq.~\eqref{gnl} with the symmetric choice $g^{(l)}=1/\sqrt{2}$, one has} $g_n^{(A)}=\delta_{n,\pm 1}/2$ and $g_n^{(B)}= e^{\mp i\phi}\delta_{n,\pm 1}/2$ and Eq.~\eqref{eq:p_final} becomes
\be
\label{eq:p_singleharmonics}
P =  - \Omega \int_{-\infty}^{+\infty} \frac{\d\omega}{4\pi m} {\cal J}_1(\omega+\Omega) N(\omega,\Omega){ \chi}_{{\rm eff}}''(\omega)~,
\ee
with
\be
\label{eq:chieff}
\!\!\!\!\chi_{{\rm eff}}(\omega)= {\chi_2^{(A,A)}}(\omega)+ {\chi_2^{(B,B)}}(\omega) + 2 \cos(\phi){\chi_2^{(A,B)}}(\omega),
\ee
an effective response function that explicitly depends on the phase $\phi$, governing the constructive/destructive interference induced by the non-diagonal term $\chi_2^{(A,B)}$.
Looking for the \red{maximum output power, with these two monochromatic drives,} it is easy to see that only two phase values are relevant, \ie $\phi =0$ or $\phi=\pi$.\red{Indeed, for $\phi = 0$, $\chi''_{{\rm eff}}(\omega)$ is peaked around $\omega \sim \omega_l$, while for $\phi=\pi$ it is peaked around $\omega\sim \bar{\omega}$.}

 To appreciate the effects of bath-mediated correlations, the output power produced by the joint configuration of Fig.~\ref{fig:1}(a) should be compared to the one obtained in the independent configuration of Fig.~\ref{fig:1}(b). In the latter case $P$ is given by an expression analogous to Eq.~\eqref{eq:p_singleharmonics} where $\chi_{{\rm eff}}(\omega)\to \chi_{{\rm eff,\rm ind}}(\omega)=-\sum_{l=A,B}(\omega^2-\omega_l^2 + i \gamma_2 \omega)^{-1}$.

In the limit of small damping strength $\gamma_2\to 0$, the response function ${\bm \chi}_2(\omega)$ reduces to the one above and therefore no \red{collective} effects are expected to show up. However we know that the response functions qualitatively change while increasing the damping strength. To see these effects in Fig.~\ref{fig:density}(a) we consider the case of a moderate damping strength $\gamma_2=0.1\omega_A$. Here, the engine output power is reported considering the temperature configuration~\cite{note_temp} $T_1=0.6\omega_A$, $T_2=0.4\omega_A$, and the representative value $\omega_B=0.6\omega_A$. The density plot in the $\omega_1-\Omega$ plane shows the working regions of the dynamical heat engine, where $P<0$. From the figure it is clear that these regions follow two distinct sectors that correspond to the lines $\omega_1=\omega_l+\Omega$ with $l=A,B$.  This is due to the Lorentzian spectral density of Eq.~\eqref{eq:lorspd} acting as an effective filter. For the representative value $\omega_B=0.6\omega_A$ the \red{maximum} output power is obtained for $\phi=0$. Already at this moderate damping $\gamma_2=0.1\omega_A$ a \red{dissipation-induced} benefit in the output power starts to emerge. This can be seen looking at the smaller values in the inset of Fig.~\ref{fig:density}(a), where the power of the independent case of Fig.~\ref{fig:1}(b) is reported.

Marked signatures of bath-mediated correlations shows up in the strong dissipation regime ($\gamma_2\gg \omega_{A,B}$), when full frequency locking is established, as reported in Fig.~\ref{fig:density}(b) with $\gamma_2=100\omega_A$.
Strikingly, the joint configuration results in a wide and sizeable working regime for the dynamical heat engine. Moreover, differently from Panel (a), the working region is now concentrated along a single line, \ie $\omega_1=\bar{\omega}+\Omega$ that extends over a wider region in the $\omega_1-\Omega$ plane. The fact that only a single region now appears is consistent with the frequency locking mechanism, with the maximum power obtained for $\phi=\pi$. In addition, comparing the moderate and strong damping situation, one can note a large increase in the output power magnitude observed for this parameter choice. The behaviour of the power as a function of the damping strength $\gamma_2$, both for the joint and the independent configuration,  is  analyzed in Fig.~\ref{fig:density}(c), where the maximum output power is reported. It is clear that above a certain critical value of $\gamma_2$ the independent configuration is fully overdamped and ceases to work as a heat engine, while on the contrary the joint configuration exhibits a solid and stable performance. Finally, in Fig.~\ref{fig:density}(d) we have reported the maximum output power of the joint case for the two opposite regimes of very weak ($\gamma_2\to 0$) and ultra--strong ($\gamma_2\to \infty$) damping, whose behaviours can be obtained in analytic form (see App.~\ref{app_d}). In the former, weak damping regime, one finds a phase--independent, completely uncorrelated power. In the latter, instead,  $\gamma_2\to\infty$ regime, the dependence on the phase $\phi$ is crucial: only for $\phi=\pi$ one obtains $P<0$. Figure~\ref{fig:density}(d) shows that a wide region of parameters exists where the strong dissipation regime can even outperform over its weak counterpart, demonstrating that frequency locking can be the optimal working point to benefit from \red{collective} effects.

\subsection{Pareto optimal performances}
Here, we generalize the analysis of the \red{dissipation-induced collective} effects on the performance of quantum thermal machine by (i) performing a functional optimization over arbitrary periodic driving functions $g_1^{(l)}(t)$, and (ii) deriving the Pareto front, \ie the collection of driving functions that are Pareto--optimal tradeoffs between power and efficiency $\eta = -P/J_1$~\cite{seoane2016}. A Pareto--optimal cycle is one such that it is not possible to further improve the power or efficiency, without sacrificing the other one. The Pareto front is then defined as the collection of $(\eta,-P)$ points of all Pareto--optimal cycles, which in general will include the maximum power case, the maximum efficiency case, and intermediate tradeoffs.
{Note that if a cycle is on the Pareto front of $(\eta,-P)$, it is also on the Pareto front of $(\sigma,-P)$, \ie it is also Pareto--optimal between high power and low entropy production. Therefore, we search for the Pareto front in $(\sigma,-P)$, and then transform these points to $(\eta,-P)$ removing the non--Pareto--optimal ones.
\red{We determine the $(\sigma,-P)$ Pareto front of the dynamical heat engine with respect to the driving coefficient $g_n^{(l)}$ expressing both thermodynamic quantities as $-P(\{g_n^{(l)}\})$  and $\sigma(\{g_n^{(l)}\})$.}
\red{As shown in the SI without loss of generality we can assume the $g_n^{(l)}$ coefficients to be real. We then consider a collection of fixed values of the entropy production rate $\{\sigma_i\}$, and for each one, we repeat the following optimization problem}
\red{
\be
-P_i = \max_{\{g_n^{(l)}\}}{\left[-P(\{g_n^{(l)}\})\right]},
\label{eq:pareto_problem}
\ee
subject to $\sigma(\{g_n^{(l)}\}) = \sigma_i$ and the two constraints in Eq.~\eqref{gnl}. }
\red{The Pareto front is then given by all points $\{(\sigma_i,-P_i) \}$. Using Parseval's theorem, the two constraints in Eq.~\eqref{gnl} are equivalent to bounding the time average of $|g_1^{(l)}(t)|^2$. This allows us to control the driving strength to both QHOs, ensuring that the coupling to bath 1 remains in the weak-coupling regime.}

\red{We solve the optimization problem in Eq.~(\ref{eq:pareto_problem}) numerically, using gradient-based optimization techniques, considering a set of $5.000$ discrete frequencies, \ie $10.000$ parameters $g_n^{(l)}$ with $n>0$ (negative frequency coefficients must be the same as positive ones to guarantee that $g^{(l)}(t)$ is real).}\red{ We then use automatic differentiation and the ADAM algorithm~\cite{kingma2014}, implemented in the PyTorch package~\cite{paszke2019}, to perform a gradient descent optimization starting from a random guess of the driving parameters.
 A modification of the Lagrange multipliers technique suitable for a gradient descent approach~\cite{platt1987} is used to enforce the entropy constraint in Eq.~(\ref{eq:pareto_problem}), while the two constraints in Eq.~\eqref{gnl}  are exactly imposed renormalizing the coefficients (see SI for details).} 
  \begin{figure}[ht] 
    	\centering
    	\includegraphics[width=1.\linewidth]{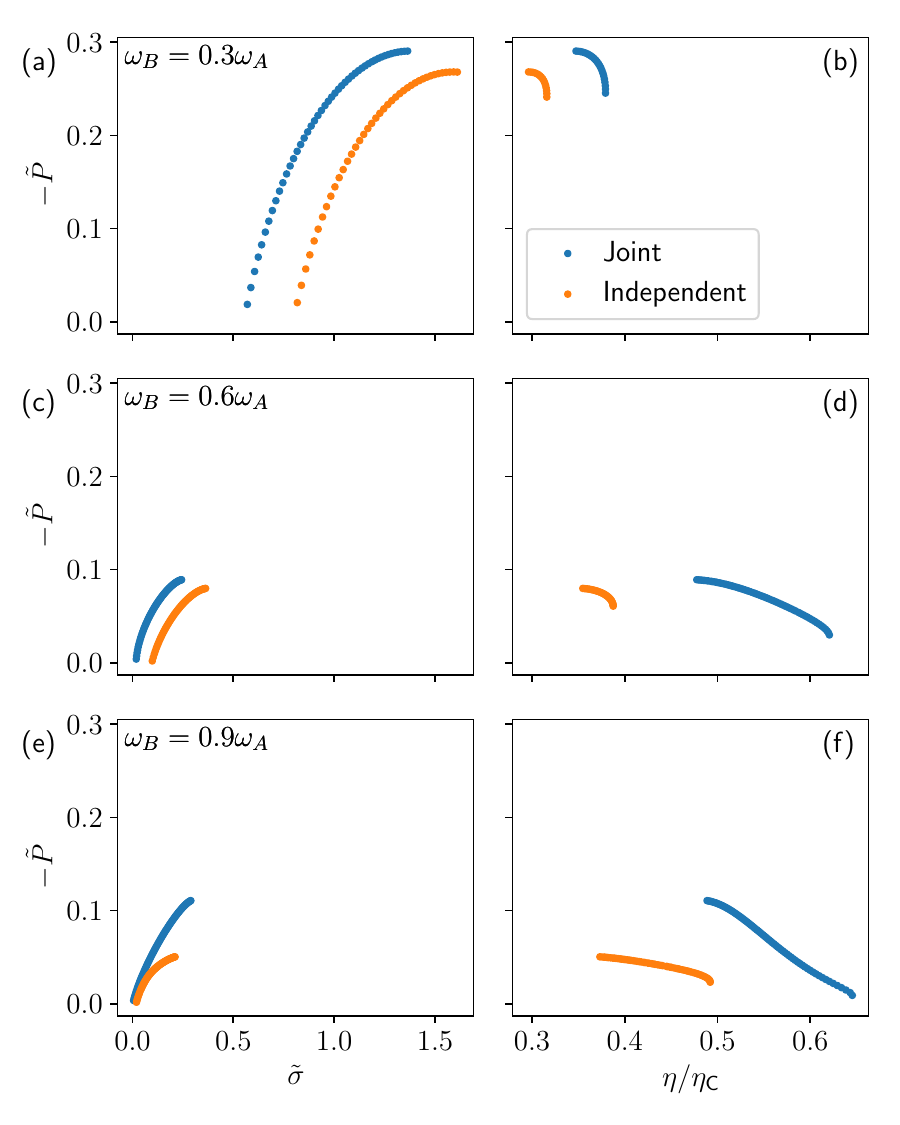}
    	\caption{Comparison between the joint (blue dots) and independent (orange dots) Pareto front in the moderate damping case, \ie $\gamma_2=0.1\omega_A$. The left column reports the Pareto front in the $(\tilde{\sigma},-\tilde{P} )$ space, and the right column reports the same points in the $(\eta/\eta_\text{C},-\tilde{P})$, where $\eta_\text{C}=1-T_2/T_1$ is the Carnot efficiency. Each row corresponds to a different value of $\omega_B/\omega_A$: (a,b) correspond to $0.3$, (c,d) to $0.6$ and (e,f) to $0.9$. For each $\omega_B$, the value of $\omega_1$ is fixed to the one that yields maximum power as in Fig.~\ref{fig:density}(a). The driving magnitudes are fixed to $|g^{(l)}|^2\equiv\sum_n |g_n^{(l)}|^2=0.5$, which are consistent with the monochromatic driving. All other system parameters are chosen as in Fig.~\ref{fig:density}. The numerical calculations are performed optimizing over $5000$ evenly spaced frequencies, for each QHO, in the $[0, 0.5 \omega_A]$ interval. All plots report dimensionless quantities, \ie $\tilde{P}= P\omega^2_A/d_1$ and $\tilde{\sigma}=\sigma/\omega_A$.
}
\label{fig:pareto01}
\end{figure}
In Fig.~\ref{fig:pareto01} we report the results for the moderate damping case ($\gamma_2=0.1\omega_A$) for three representative values of $\omega_B$. In all cases, the value of $\omega_1$ has been chosen as the one that yields the maximum output power. 

Notably, in all cases, the entire Pareto front of the bath-mediated situation in the joint case is strictly better than that of the independent configuration, \ie for all points along the Pareto front of the independent case, there is at least one point in the joint case that yields higher power \textit{and} higher efficiency. Furthermore, not only is the maximum power higher, but especially the efficiency of the \red{joint} case is enhanced along the entire Pareto front, reaching values that are twice as large in the $\omega_B/\omega_A=0.6, 0.9$ cases -- see panels (c-f) of Fig.~\ref{fig:pareto01}. Interestingly, as we move from $\omega_B=0.3\omega_A$ to $\omega_B=0.9\omega_A$, the Pareto fronts move from a region of high power and low efficiency -- upper left corner in panels (b,d,f) --  to a region of lower power but higher efficiency -- lower right corner in panels (b,d,f).

Again, the effect of the bath-mediated interaction between the QHOs is clearly visible, even in the moderate damping case, comparing the optimal driving in the joint and independent case. Indeed, in the latter, the optimal driving consists of applying two different frequencies to each QHO: intuitively, this is expected, since each QHO has a different characteristic frequency. However, in the joint case the optimal driving turns out to be monochromatic along the entire Pareto front, for all explored values of $\omega_B$ except for $\omega_B=0.6 \omega_A$, where two frequencies become optimal when the power is lower than $|\tilde{P}| \sim 0.086$ (see SI).
 
 \begin{figure}[ht] 
    	\centering
    	\includegraphics[width=1.\linewidth]{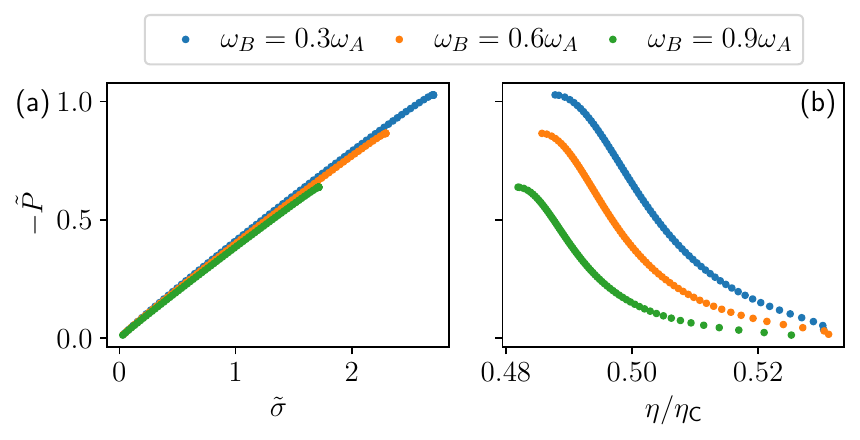}
    	\caption{Pareto front of the joint case in the ultra--strong damping regime, \ie $\gamma_2=100\omega_A$. As in Fig.~\ref{fig:pareto01}, (a) reports the Pareto front in the $(\tilde{\sigma},-\tilde{P})$ space, and (b) reports the same points in the $(\eta/\eta_\text{C},-\tilde{P})$. Each curve corresponds to different values of $\omega_B/\omega_A$ as shown in the legend. 
The driving magnitudes are fixed to $|g^{(l)}|^2=0.5$, which are consistent with the monocromatic driving. All other system parameters are chosen as in Fig.~\ref{fig:density}. The numerical calculations are performed optimizing over $5000$ evenly spaced frequencies, for each QHO, in the $[0, 0.5 \omega_A]$ interval.}
\label{fig:pareto100}
\end{figure}
In Fig.~\ref{fig:pareto100} we report the results of the joint case for the strong dissipation regime at $\gamma_2/\omega_A=100$. As in Fig.~\ref{fig:pareto01}, we set $\omega_1$ to the value that yielded maximum power in the corresponding $\omega_1-\Omega$ plane and the left and right Panels correspond, respectively, to the Pareto front in the $(\sigma,-P)$ and $(\eta,-P)$ space. The three curves correspond to the different values of $\omega_B$ reported in the legend.

Remarkably, the strong dissipation regime displays a high-performance Pareto front, reaching values of the power that are roughly three times larger than in the moderate damping regime, while operating at a high efficiency $\eta\sim 0.5\eta_\text{C}$.
In addition, the optimal driving along the entire Pareto front always consists of a monochromatic drive (see SI).

\subsection{Measuring entanglement via average work}
Before closing, we report another important result of the strong damping regime: we will demonstrate a direct prescription to assess  the degree of entanglement for the WM  from  measurements of average works. The quantifier we use for detecting quantum correlations is the so-called logarithmic negativity~\cite{simon2000, serafini2004, rebecca, henkel21, paz09} $E_n$ (see SI).
Here we discuss a possible pathway to measure the degree of entanglement for a quantum system via thermodynamic observables. 

We recall that the logarithmic negativity is defined as~\cite{simon2000, plenio_prl, paz09} 
\be
\label{main:en_def}
E_n \equiv {\rm Max} \big[0, - \log(2\tilde{\nu})\big]~,
\ee
where $\tilde{\nu}$ is the so-called symplectic eigenvalue of the partial transposed density matrix. Strictly positive values of $E_n$ are  a fingerprint of entanglement. To be in this regime, $\tilde{\nu} < 1/2$ is required and this implies a constraint on temperatures: only for $T<T_{{\rm c}}$, with 
$T_{{\rm c}}$ a critical temperature, the WM will be entangled.
The behaviour of $T_{{\rm c}}$ is shown in Fig.~\ref{fig:Tc} as a function of $\omega_B$ for different damping strengths $\gamma_2$. As one can see $T_{{\rm c}}$ tends to saturate to the value $T^*$ for $\gamma_2\to\infty$.
\begin{figure}[ht] 
    	\centering
        	\includegraphics[width=0.85\linewidth]{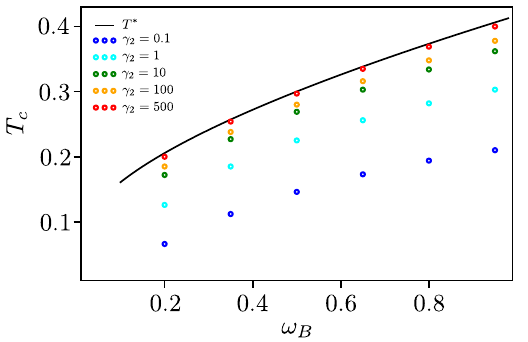}
 \caption{Plot of the critical temperature $T_{{\rm c}}$ below which non-zero entanglement in the WM is expected as a function of $\omega_B$ for different values of $\gamma_2$ (see legend). The black solid line is the plot of the asymptotic critical temperature $T^*$ for $\gamma_2\to\infty$. All temperatures are in units of $\omega_A$.
\label{fig:Tc}}
\end{figure}
Since $T^*$ is always the  upper bound with respect to all other critical temperatures we focus on the best working point at ultra--strong damping. In this regime we found a closed analytic expression for $\tilde{\nu}$ (see SI) given by 
\red{
\be
\tilde{\nu}^2=\frac{\bar{\omega}^3 T_2 \coth^2\left(\frac{\bar{\omega}}{2T_2}\right)}{2 (\bar{\omega}^4-\Delta^4)}\frac{1}{\coth\left(\frac{\bar{\omega}}{2T_2}\right)+\frac{2T_2\Delta^4}{ \bar{\omega} (\bar{\omega}^4-\Delta^4)}}\,.
\label{eq:nutilde2_final}
\ee
}
Using the above expression we  obtain the critical temperature $T^*$  shown as a black curve in Fig.~\ref{fig:Tc}.

We will now demonstrate a direct link between $\tilde{\nu}^2$ in Eq.~\eqref{eq:nutilde2_final} and a combination of works $W$ (obtained from  the average power as $W=(2\pi/\Omega)P(\Omega)$) evaluated at different working points of the machine.
\begin{figure}[ht] 
    	\centering
    	    	\includegraphics[width=0.75\linewidth]{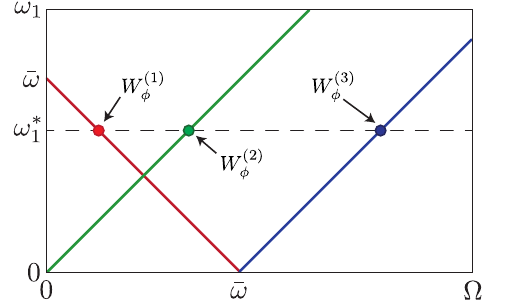}
  \caption{Sketch of the three working points in the $\omega_1-\Omega$ plane where we evaluate the average work
$W_\phi^{(1,2,3)}$. The horizontal dotted line highlights the representative value for $\omega_1=\omega_1^*$ around which  the spectral density ${\cal J}_1(\omega)$ is peaked.
\label{fig:lines}}
\end{figure}

As shown in the sketch of Fig.~\ref{fig:lines} we select three working points in the $\omega_1-\Omega$ plane, obtained at the intersection  between the horizontal dotted line at a fixed $\omega_1=\omega_1^*<\bar{\omega}$ and 
three lines: $\omega_1 = \bar{\omega}-\Omega$ (red), $\omega_1 =\Omega$  (green), $\omega_1 = -\bar{\omega} + \Omega$ (blue). In addition we consider a peaked spectral function ${\cal J}_1(\omega)$, acting as a sharp filter.

Using monochromatic drives, in the ultra--strong damping regime the average work crucially depends on $\phi$. We now inspect the following combination of these average works: $\Delta W_\phi=W_\phi^{(2)}+W_\phi^{(3)}-W_\phi^{(1)}$. The result (see App.~\ref{app_d}) is
\red{\beq
\Delta W_{\phi=\pi} &=& \frac{\pi}{m\bar{\omega}} \left[\coth\left(\frac{\bar{\omega}}{2T_2}\right)+ \frac{2T_2\Delta^4}{ \bar{\omega}(\bar{\omega}^4-\Delta^4)}\right]{\cal J}_1(\omega_1^*)\,\nonumber\\
\Delta W_{\phi=0} &=& \frac{2\pi T_2}{m}\frac{\bar{\omega}^2}{\bar{\omega}^4-\Delta^4}{\cal J}_1(\omega_1^*).
\label{DeltaW}
\eeq
}
Comparing these results with $\tilde{\nu}^2$ in Eq.~\eqref{eq:nutilde2_final} we arrive at the important identity 
\be
\tilde{\nu}^2= \frac{\coth^2\left(\frac{\bar{\omega}}{2T_2}\right)}{4}\frac{\Delta W_{\phi=0}}{\Delta W_{\phi=\pi}}~.\label{eq:link}
\ee
This expression represents a direct link between entanglement and a function of the average works computed for specific driving protocols. Equation \ref{eq:link} is universal and does not depend on the particular working regime of the thermal machine, provided that ultra--strong damping has been reached.
\section{Discussion}
We have shown that dissipation can trigger a \red{collective} advantage for a quantum heat engine made of two non-interacting quantum harmonic oscillators connected to common heat baths, with the couplings to one of them periodically driven. \red{This} advantage is rooted in the non-trivial correlations between the oscillators mediated by the baths. Of particular interest is the regime of strong dissipation, where  two independent single-oscillator engines working in parallel can not deliver any power, whereas with common baths the engine shows high performances. 
The fact that the strong dissipation is a useful resource might seem surprising, but we have explained this result in terms of the appearance of a (weakly damped) frequency- and phase-locked mode, with the two oscillators moving with a \red{common} frequency and oscillating in phase opposition.
The claim of \red{collective} advantage has been corroborated by the optimization over generic periodic driving protocols, building the full Pareto fronts and thus providing the optimal tradeoffs between power and efficiency. As a final outcome of our work, we have found a precise prescription in terms of thermodynamic quantities, such as average works, which allows to assess the degree of entanglement of the whole quantum system.

Our results open up several perspectives. First of all, it would be interesting to investigate whether the dissipation-induced \red{collective} advantage can be extended to $N>2$ oscillators and possibly to establish a link between \red{collective} advantage and multipartite quantum correlations. Secondly, it would be interesting to consider different, non linear working media, in particular hybrid oscillator-qubit systems of particular interest for quantum computing and more generally for 
quantum technologies. Thirdly, our Pareto front analysis paves the way to the use of machine learning tools for non-Markovian quantum thermodynamics processes. 
\appendix
\section{Thermodynamic observables}
\label{app_b}
The average power of Eq.~\eqref{averagepower} is associated to the time--dependent power operator, induced by the time varying coupling coefficients $g_1^{(l)}(t)$ that modulate the $\nu=1$ WM/bath coupling. Its expression is
\begin{align}
\label{power}
\!\!\!{P}(t) &= \frac{\partial H^{(t)}_{{\rm int},1}(t)}{\partial t} = \sum_{l=A,B} \sum_{k=1}^{+\infty} \bigg[
- \dot{g}_1^{(l)}(t) c_{k,1} x_l(t) X_{k,1}(t)\nonumber\\
&\quad
+g_1^{(l)}(t) \dot{g}_1^{(l)}(t) \frac{c_{k,1}^2}{m_{k,1}\omega_{k,1}^2} x_{l}^2(t)\nonumber\\ 
&\quad
+\partial_t [g_1^{(l)}(t) g_1^{(\bar{l})}(t)] \frac{ c_{k,1}^2  }{2 m_{k,1}\omega_{k,1}^2} x_{l}(t) x_{\bar{l}}(t) \bigg]\,.
\end{align}
The average heat currents in Eq.~\eqref{averagecurrent} instead describe the heat flows from or toward the $\nu$-th reservoirs, and depend on the operators $J_{\nu}(t)$ given by
\be
\!\!\!\!\!\!J_\nu (t) = -\dot{H}_{\nu} (t)= - \!\!\!\sum_{l = A,B}
g_{\nu}^{(l)}(t) x_l(t) \sum_{k=1}^{+\infty}c_{k,\nu}\dot{X}_{k,\nu}(t)\,.
\label{eq:Jnu_def}
\ee
It is clear that these quantities depend on the WM and baths position operators, that obey a set of coupled equations of motion (see SI).
The bath position operators can be expressed exactly as
\begin{align}
X_{k,\nu} (t) & = \xi_{k,\nu} (t) + \sum_{l=A,B} \frac{c_{k,\nu}}{m_{k,\nu} \omega_{k,\nu}} \int_{t_0}^{t} \d s \, g_{\nu}^{(l)}(s) x_l(s)\nonumber\\
&\quad \times \sin[\omega_{k,\nu}(t-s)],
\label{eq:X_R_sol_EoM1}
\end{align}
where 
\begin{align}
\xi_{k,\nu}(t)\equiv&  X_{k,\nu} (t_0) \cos [\omega_{k,\nu}(t-t_0)] \nonumber\\
& + \frac{P_{k,\nu}(t_0)}{m_{k,\nu} \omega_{k,\nu}}\sin[\omega_{k,\nu}(t-t_0)].
\label{eq:xi_knu_def}
\end{align}
The corresponding fluctuating force is
\begin{equation}
\xi_\nu(t)\equiv\sum_{k=1}^{+\infty} c_{k,\nu}\xi_{k,\nu}(t),
\label{eq:xi_nu_def1}
\end{equation}
with zero quantum average $\langle \xi_\nu(t)\rangle =0$ and correlator~\cite{weiss} 
\begin{align}
\label{xixi}
&\langle \xi_\nu(t) {\xi}_{\nu'}(t') \rangle = \delta_{\nu,\nu'}\int_0^\infty \frac{\d\omega}{\pi} {\cal J}_\nu (\omega)  \nonumber\\
& \times \left[ \coth\left(\frac{\omega}{2 T_\nu}\right) \cos[\omega (t-t')] - i \sin[\omega (t-t')]\right].
\end{align}
Looking at Eqs.~\eqref{power},~\eqref{eq:Jnu_def}, and~\eqref{eq:X_R_sol_EoM1}, the behaviour of these thermodynamic quantities is determined by the dynamics of $x_A(t)$ and $x_B(t)$.

\section{Time evolution of QHO operators and response functions}
\label{app_c}
The WM is weakly coupled to the $\nu=1$ reservoir. Therefore, in the main text we have considered a perturbative expansion under which the position operator of the $l$-th QHO can be then written as
\begin{equation}
x_l(t)= x_l^{(0)}(t) + \Delta x_l(t),~\label{eq:x_qho_pert} 
\end{equation}
where $x_l^{(0)}(t)$ evolves under the unperturbed Hamiltonian and is influenced only by the coupling with the static reservoir $\nu=2$.
The perturbative correction due to $H_{{\mathrm{int}},1}^{(t)}$ is (see SI)
\begin{equation}
\!\!\!\Delta x_l(t)=\!-i\!\!\sum_{l'=A,B}\int_{t_0}^t \d s  g_1^{(l')}(s) \xi_1(s)\big[x_{l'}^{(0)}(s),x_{l}^{(0)}(t)\big], 
\label{eq:corrxl0}
\end{equation}
with $\xi_1(t)$ defined in Eq.~\eqref{eq:xi_nu_def1}.
From this expression, it is clear that to evaluate Eqs.~\eqref{averagepower}--\eqref{averagecurrent} via Eqs.~\eqref{power}--\eqref{eq:Jnu_def}, one needs the fluctuating force correlators \cite{weiss} $\langle \xi_1(t)\xi_1(t')\rangle$, given in Eq.~\eqref{xixi}, and $\langle x_l^{(0)}(t)x_{l'}^{(0)}(t')\rangle$, which depends only on the static (unperturbed) bath $\nu=2$. The latter are needed to evaluate the response functions $\chi_{2}^{(ll')}(t)$ defined in Eq.~\eqref{eq:responsefunction}. To obtain its expression and study its properties, we begin considering the coordinates $x^{(0)}_l(t)$ of the WM, which satisfy a set  of coupled Langevin equations (see also SI). These equations can be conveniently written in Fourier space
\begin{align}
[-\omega^2+ \omega_A^2 &- i \omega \gamma_2(\omega)]x^{(0)}_A(\omega)  \nonumber\\
& - i \omega  \gamma_2(\omega)x^{(0)}_B(\omega) =\frac{\xi_2(\omega)}{m}~,\nonumber \\
[-\omega^2 + \omega_B^2 &- i \omega \gamma_2(\omega)] x^{(0)}_B(\omega) \nonumber\\
& -i\omega   \gamma_2(\omega)x^{(0)}_A(\omega)=\frac{\xi_2(\omega)}{m}\,.
\label{langevin1}
\end{align} 
They depend on the noise term $\xi_2(\omega)=\int_{-\infty}^{\infty}\d t e^{i\omega t}\xi_2(t)$ with $\xi_2(t)$  in Eq.~(\ref{eq:xi_nu_def1}) and on the damping kernel in Fourier space $\gamma_2(\omega)$, defined in terms of the time kernel $\gamma_2(t)$ (see SI) as
 \be
 \gamma_2(\omega)=\int_{-\infty}^{+\infty}\d t e^{i\omega t}\theta(t)\gamma_2(t)~.
 \ee
To understand the physics of two QHOs with bath-mediated interactions, we study in Eq.~\eqref{langevin1} the intrinsic excitations of the normal modes in the absence of the noise term (\ie setting $\xi_2(\omega)=0$).
Following the standard procedure~\cite{weiss}, they are given by the zeros of $D(\omega)$ in Eq.~\eqref{eq:denominator2}, which is the determinant of the coefficient matrix of Eq.~(\ref{langevin1}).These zeros,  which are four in our case: $z_j= z_j' + i z_j''$, determine the dynamics of $x^{(0)}_{A,B}(t)$ at finite times.
In particular, their real parts give the possible frequencies of oscillations, while the imaginary ones describe their damping.  
Their explicit form will depend on the shape of the damping kernel $\gamma_2(\omega)$ and, eventually, on the spectral density of the static bath $\nu=2$. Here, we consider a Ohmic spectral density with a Drude cut--off ${\cal J}_2(\omega)=m\omega\gamma_2/(1+\omega^2/\omega_c^2)$ which leads to $\gamma_2'(\omega)={\cal J}_2(\omega)/m\omega$ and $\gamma_2''(\omega)={\cal J}_2(\omega)/m\omega_c$~\cite{weiss}, with $\omega_c$ the high energy cut-off. When not necessary we will consider the latter as the highest energy of the system ($\omega_c\to\infty$) with then 
 ${\cal J}_2(\omega)=m\omega\gamma_2$, and a constant and real $\gamma_2(\omega)=\gamma_2$.

\begin{figure}
 \centering
 \includegraphics[width=0.8\linewidth]{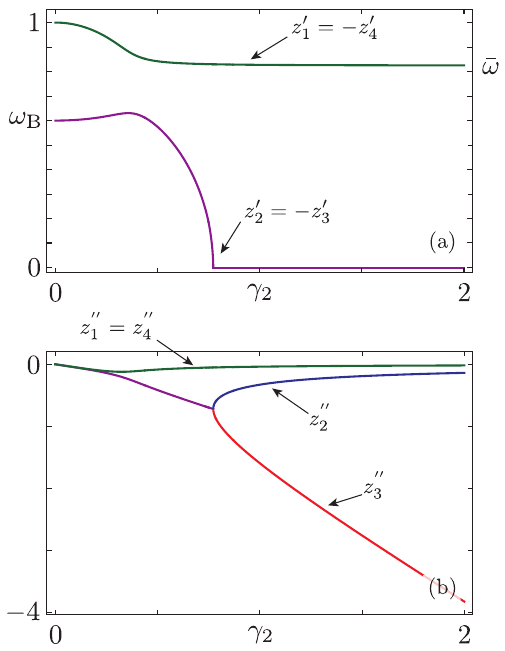}
        \caption{Plot of (a) the real part  $z'_j$ and (b) the imaginary part $z_j''$ of the four zeros of $D(\omega)$ for the case of a Ohmic spectral density for the $\nu=2$ bath in the limit of large cut--off $\omega_c\to\infty$, as a function of the damping strength $\gamma_2$ for representative parameter values $\omega_A=1$ and $\omega_B=0.6$. All quantities are in units of $\omega_A$. 
\label{fig:poles}}
\end{figure}
In Fig.~\ref{fig:poles} we report the real -- Panel (a) -- and imaginary -- Panel (b) -- parts of the four zeros  as a function of the damping strength $\gamma_2$.  Looking at Fig.~\ref{fig:poles}(a) it is easy to see that for $\gamma_2\ll \omega_{A,B}$ the system behaves as two independent QHOs with characteristic frequencies $\omega_A$ and $\omega_B$. On the other hand, at large damping \red{$\gamma_2 \gg \Delta^4/\bar{\omega}^3$ (with $\bar{\omega}$ in Eq.~\eqref{eq:hybrid} and $\Delta^2 =(\omega_A^2-\omega_B^2)/2$)}  two modes are frequency locked to the \red{common frequency $\bar{\omega}$} , with $z_1'=-z_4'\to \bar{\omega}$, while the other two are overdamped, namely $z_2'=-z_3'= 0$. 

Considering now the imaginary parts, shown in Fig.~\ref{fig:poles}(b) we see that for small damping they all are $\propto \gamma_2$, and acquire different behaviours at larger $\gamma_2$. In particular the zeros that tend to a finite frequency $z_1'=-z_4'\to \bar{\omega}$, have imaginary parts that tend to vanish with increasing $\gamma_2$. On the other hand, the overdamped ones  $z_2$ and $z_3$ possess imaginary contributions that run together until a critical value and bifurcate after it with opposite behaviour. 

In the regime of very weak $\gamma_2\ll \omega_{A,B}$ we obtain 
\be
z_{1,4}= \pm \omega_A- i \frac{\gamma_2}{2};\quad z_{2,3}= \pm \omega_B - i\frac{\gamma_2}{2},
\label{eq:smallgamma}
\ee
while at ultra--strong damping \red{$\gamma_2\gg \Delta^4/\bar{\omega}^3$} we have
\red{
\be
\!\!z_{1,4}= \pm\bar{\omega} - i \frac{\Delta^4}{4\gamma_2 \bar{\omega}^2};\,
z_2= -i\frac{(\bar{\omega}^4-\Delta^4)}{2\gamma_2\bar{\omega}^2};\, z_3 = -2i\gamma_2,
\label{eq:largegamma}
\ee
}
in full agreement with the behaviour reported  in Fig.~\ref{fig:poles}. 
Notice that the zeros that survive at the common frequency $\bar{\omega}$ are  very stable, \ie with a very small imaginary part $\propto 1/\gamma_2$.

The above results demonstrate that  at strong damping, the two QHOs become {\it frequency locked}  oscillating, at finite time, with a common frequency $\bar{\omega}$. Moreover, they are also {\it phase locked} in anti--phase (with relative phase $\pi$), as one can see from the relation of the homogeneous solution of Eq.~(\ref{langevin1}) with $x_A(\omega)=-x_B(\omega)$ at large $\gamma_2$. 
All these regimes are reflected on the retarded response function. Indeed, switching on the noise $\xi_2(\omega)$, the system~(\ref{langevin1}) has the long time compact solution 
${\bf x}^\dag ={\bm \chi}_2 \cdot {\bm \xi }^\dag/m$, where ${\bf x} = (x^{(0)}_A(\omega) , x^{(0)}_B(\omega))$ is  the two-component vector of the positions of the oscillators, ${\bm \xi}=\xi_2(\omega) (1,1)$ is the noise vector and
${\bm \chi}_2(\omega)$ is the two-by-two response-function matrix, inverse of the coefficient matrix. Its elements are the Fourier transform of Eq.~\eqref{eq:responsefunction} and are given by
\be
\label{eq:chill}
\!\!\!\!\!\chi_2^{(l l)}(\omega)\!=\!\frac{-[\omega^2-\omega_{\bar{l}}^2 + i \omega \gamma_2(\omega)]}{D(\omega)};\,
\chi_2^{(l \bar{l})}(\omega)\!=\!\frac{i\omega\gamma_2(\omega)}{D(\omega)}.
\ee
The behaviour of the response function is connected to the already discussed normal modes. We are particularly interested in the imaginary part of such response function.

At very weak damping ($\gamma_2\ll\omega_{A,B}$), ${\bm \chi}_2''(\omega)$  is exactly the one of two independent QHOs 
(see also Eq.~(\ref{eq:smallgamma})):
\be
{\chi_2^{(l l')}}''(\omega)=\delta_{l,l'}\frac{i\omega\gamma_2}{(\omega^2-\omega_l^2)^2 +  \omega^2\gamma^2_2}.
\label{chigamma0}
\ee
In the opposite regime (\red{$\gamma_2\gg \Delta^4/\bar{\omega}^3$}) from Eq.~(\ref{eq:largegamma}) we obtain
\red{
\begin{align}
\!\!\!\!&{\chi_2^{(l l')}}''(\omega)= \Big[
(1-\delta_{l,l'})+\frac{\omega_{\bar{l}}^2}{\omega_l^2}\delta_{l,l'}\Big]\frac{\omega}{\omega_l^2+\omega^2_{\bar{l}}}
\frac{|z_2''|}{\omega^2 + |z_2''|^2} \nonumber\\
\!\!\!\!&+\!\frac{\omega (-1)^{1-\delta_{l,l'}}}{2(\omega_l^2+\omega^2_{\bar{l}})}\!\sum_{p=\pm}\!\frac{|z_1''|}{\left(\omega\!+\!p\sqrt{(\omega^2_l+\omega^2_{\bar{l}})/2}\right)^2 \!\!\!+ |z_1''|^2}.
\label{eq:llstrong}
\end{align}
}
Here, since $z_1''$ and $z_2''$ scale as $1/\gamma_2$ -- see Eq.~\eqref{eq:largegamma} -- both the diagonal and the off diagonal response functions are dominated by contributions peaked around \red{$\pm \bar{\omega}=\pm \sqrt{(\omega_A^2+\omega_B^2)/2}$}. Indeed in this regime, the WM 
becomes effectively {\it frequency locked} to a unique common frequency, \ie $\bar{\omega}$. In addition, the  phase locking at $\pi$, discussed above, is here reflected in the property ${\chi_2^{(l \bar{l})}}''(\omega)\to -{\chi_2^{(l l)}}''(\omega)$. Note that the first term in Eq.~\eqref{eq:llstrong} $\propto \omega$, although negligible with respect to the second one, can give a finite dissipative contributions to the power (see below).

For the sake of comparison, we quote the behaviour  associated to the independent configuration (see Fig.~\ref{fig:1}(b)). In this case the response funtion is diagonal and given by~\cite{cavaliere_prr} $\chi_{2,\rm ind}^{(l l')}(\omega)=-\delta_{l,l'}/[\omega^2-\omega^2_l + i \omega\gamma_2]$. It is then straighforward to see that at very weak damping the normal modes  are the same as in the joint case in Eq.~(\ref{eq:smallgamma}), while for strong damping they are given by $z_{1,\rm ind}= -i\omega_A^2/\gamma_2$, $z_{2,\rm ind}= -i\omega_B^2/\gamma_2$ and $ z_{3,4,\rm ind}= -i\gamma_2$. As we can see, in the latter case,  all zeros are purely imaginary, implying always an overdamped regime at strong damping. This behaviour is in sharp contrast with the case discussed above of the joint configuration.
\section{Average power for weak and strong damping}
\label{app_d}
To obtain expressions for the average power $P$ in the weak and in the strong damping regime, we start from Eq.~\eqref{eq:p_singleharmonics} and the expressions for ${\chi_2^{(l \bar{l})}}''(\omega)$ quoted above.\\ 
In the very weak damping regime $\gamma_2\ll\omega_{A,B}$ the average power is phase--independent and using Eq.~\eqref{chigamma0} we get
\be
P = -\frac{\Omega}{8m}\sum_{l=A,B}\frac{1}{\omega_l}\sum_{p=\pm}p {\cal J}_1(p\omega_l+\Omega)N(p\omega_l,\Omega).
\label{eq:p_gammazero}
\ee

In the opposite regime of ultra--strong damping \red{$\gamma_2 \gg \Delta^4/\bar{\omega}^3$} we use instead Eq.~\eqref{eq:llstrong}. Here, the power depends on the phase $\phi$. When $\phi=\pi$ one finds
\be
\label{eq:p_gammainf_pi}
\!\!P_{\pi}\!=\!\frac{\Omega}{4m\bar{\omega}}\!\Big[\frac{T_2(\omega_B^2\! -\! \omega_A^2)^2}{\bar{\omega}\omega_A^2\omega_B^2}\!{\cal J}_1(\Omega)
-\!\!\sum_{p=\pm}\!\!p {\cal J}_1 (p\bar{\omega}+\Omega)N(p\bar{\omega},\Omega)\Big],
\ee
while for the case $\phi=0$ the dominant contributions of $\chi_{2}^{(l l')}(\omega)$, around the characteristic frequency $\bar{\omega}$, cancels out in $\chi_{\rm eff}(\omega)$ and only the dissipative part around $\omega\sim 0$ remains yielding
\be
\label{eq:p_gammainf_0}
P_{0}=\frac{\Omega}{2m}T_2 {\cal J}_1(\Omega)\left(\frac{1}{\omega_A^{2}} 
+\frac{1}{\omega_B^{2}}\right)>0\,,
\ee
\ie no useful power can be delivered. Importantly, we notice that the corresponding power of the independent case, $P_{\mathrm{ind}}$, coincides exactly with the above result: $P_{\mathrm{ind}}\equiv P_0$. Indeed, in this regime we have $\chi_{\mathrm{eff}}^{\mathrm{ind}}(\omega )=\sum_{l=A,B}i/[\gamma_2(\omega+i\omega_{l}^{2}/\gamma_2)]$.

From the above equations, it is easy to obtain the analytic expressions of the average work at ultra--strong damping along the lines of Fig.~\ref{fig:lines}. For $\phi=\pi$ we have
\beq
&&W^{(1)}_{\phi=\pi}=-\frac{\pi}{2m\bar{\omega}} {\cal J}_1(\omega_1^*)N(-\bar{\omega},\bar{\omega}-\omega_1^*)\nonumber \\
&&W^{(2)}_{\phi=\pi}=\frac{\pi}{2m}\frac{T_2}{\bar{\omega}^2}\frac{(\omega_B^2-\omega_A^2)^2}{\omega_A^2\omega_B^2}{\cal J}_1(\omega_1^*)\nonumber \\
&&W^{(3)}_{\phi=\pi}=\frac{\pi}{2m\bar{\omega}} {\cal J}_1(\omega_1^*)N(-\bar{\omega},\bar{\omega}+\omega_1^*)\,,
\eeq
while for $\phi=0$ we obtain 
\be
W^{(2)}_{\phi=0}=\frac{\pi T_2}{m} \left(\frac{1}{\omega_A^{2}} 
+\frac{1}{\omega_B^{2}}\right){\cal J}_1(\omega_1^*)
\ee
and $W^{(1)}_{\phi=0}, W^{(3)}_{\phi=0}\to 0$.
From the above expressions it is immediate to obtain the expressions for $\Delta W_{\phi}$ quoted in Eq.~\eqref{DeltaW}.

\section*{Data availability}
The generated numerical data presented in the plots of this paper are available from the corresponding author upon reasonable request.

\section*{Code availability}
The code used for generating data of this study are available from the corresponding author upon reasonable request.

\section*{Acknowledgements}
M.C. and M.S. acknowledge support from the project PRIN 2022 - 2022PH852L (PE3) TopoFlags - "Non reciprocal supercurrent and topological transition in hybrid Nb-InSb nanoflags" funded within the programme "PNRR Missione 4 - Componente 2 - Investimento 1.1 Fondo per il Programma Nazionale di Ricerca e Progetti di Rilevante Interesse Nazionale (PRIN)", funded by the European Union - Next Generation EU. L.R. and G.B. acknowledge financial support from the Julian Schwinger Foundation (Grant JSF-21-04-0001), from INFN through the project “QUANTUM”, and from the project PRIN 2022 - 2022XK5CPX (PE3) SoS-QuBa - "Solid State Quantum Batteries: Characterization and Optimization" funded within the programme "PNRR Missione 4 - Componente 2 - Investimento 1.1 Fondo per il Programma Nazionale di Ricerca e Progetti di Rilevante Interesse Nazionale (PRIN)", funded by the European Union - Next Generation EU. P. A. E. gratefully acknowledges funding by the Berlin Mathematics Center MATH+ (AA2-18).

\section*{Data availability}
 The generated numerical data presented in the plots of this paper are available from the corresponding author upon reasonable request.

\section*{Author contributions}
G. B. and M. S. conceived the idea. M. C., F. C., L. R., and M. S. developed the model and performed the calculations. P. A. E. studied the optimization protocol \red{with appropriate optimization techniques}.
All authors contributed to the writing of the manuscript.
\section*{Competing interests}
The authors declare no competing interests.
\section*{Additional information}
Further technical details can be found in the supplementary information.


\begin{thebibliography}{99}
\bibitem{note_ohmicbath}
\red{Note that in  the case of a strictly Ohmic spectral function ${\cal J}_1(\omega)=m\gamma_1\omega$ no working heat  engine regime is expected~\cite{cavaliere_prr}.}
\bibitem{note_temp}
We have investigated other temperature configurations in the engine working mode and no qualitative changes have been observed.\red{ Furthermore, we underline that higher values of $\gamma_1$ than the other reported would produce smaller output power magnitudes, with no qualitative changes for the collective advantage.}
\end{thebibliography}
\end{document}


\title{Supporting information of ``Dissipation-induced \red{collective} advantage in a quantum thermal machine''}
\author{Matteo Carrega}
\email{matteo.carrega@spin.cnr.it}
\affiliation{CNR-SPIN,  Via  Dodecaneso  33,  16146  Genova, Italy}
\author{Luca Razzoli}
    \affiliation{Center for Nonlinear and Complex Systems, Dipartimento di Scienza e Alta Tecnologia, Universit\`a degli Studi dell'Insubria, via Valleggio 11, 22100 Como, Italy} 
    \affiliation{Istituto Nazionale di Fisica Nucleare, Sezione di Milano, via Celoria 16, 20133 Milano, Italy}
\author{Paolo Andrea Erdman}
\affiliation{Freie Universit{\" a}t Berlin, Department of Mathematics and Computer Science, Arnimallee 6, 14195 Berlin, Germany}
\author{Fabio Cavaliere}
    \affiliation{Dipartimento di Fisica, Universit\`a di Genova, Via Dodecaneso 33, 16146 Genova, Italy} 
    \affiliation{CNR-SPIN,  Via  Dodecaneso  33,  16146  Genova, Italy}
\author{Giuliano Benenti}
    \affiliation{Center for Nonlinear and Complex Systems, Dipartimento di Scienza e Alta Tecnologia, Universit\`a degli Studi dell'Insubria, via Valleggio 11, 22100 Como, Italy} 
    \affiliation{Istituto Nazionale di Fisica Nucleare, Sezione di Milano, via Celoria 16, 20133 Milano, Italy}
    \affiliation{NEST, Istituto Nanoscienze-CNR, P.zza San Silvestro 12, I-56127 Pisa, Italy}
    \author{Maura Sassetti}
    \affiliation{Dipartimento di Fisica, Universit\`a di Genova, Via Dodecaneso 33, 16146 Genova, Italy} 
    \affiliation{CNR-SPIN,  Via  Dodecaneso  33,  16146  Genova, Italy}
\begin{abstract}
This file contains supporting materials associated to the manuscript entitled ``Dissipation-induced \red{collective} advantage in a quantum thermal machine''.
\end{abstract}
\maketitle
\section*{Supplementary Note 1: The counter-term of a dissipative two oscillators system}
\label{app:counter}
In this Section we derive the counter-term that appears in the interaction Hamiltonian $H^{(t)}_{{\rm int},\nu}$, in the presence of two  baths $\nu=1,2$.
The $\nu$-th  bath is of Caldeira--Leggett type~\cite{caldeira1983}, with free Hamiltonian
\begin{equation}
  H_\nu = \sum_{k=1}^{+\infty} \left[ \frac{P_{k,\nu}^2}{2 m_{k,\nu}} + \frac{1}{2} m_{k,\nu} \omega_{k,\nu}^2
  X_{k,\nu}^2 \right]~.
\end{equation}
The coupling between the WM and the $\nu$-bath is of the form
\be
H^{(t)}_{{\rm int},\nu}= -\Big[\sum_{l=A,B}\sum_{k=1}^{+\infty}g_\nu^{(l)}(t) c_{k,\nu}^{(l)}x_l X_{k,\nu}\Big] + H_{{\rm c.t.},\nu}.
\label{A:Hint}
\ee
Here, the last term $H_{{\rm c.t.},\nu}$ represents the so-called counter-term contribution that prevents possible renormalizations of the QHOs potentials 
$V_l(x_l)=\frac{1}{2} m \omega_l^2 x_l^2$ with $l=A,B$. Indeed, it is well known that the first term of $H^{(t)}_{{\rm int},\nu}$ introduces a renormalization of the frequencies $\omega_l$ and induces a direct coupling between the QHOs~\cite{ingold02}. 
The counter-term is then inserted in order to avoid these effects. Since the couplings between the two oscillators, $x_A$ and $x_B$, and the bath  coordinates  $X_{k,\nu}$ are linear, the general form of $H_{{\rm c.t.},\nu}$ will be 
\be 
H_{{\rm c.t.},\nu} = \lambda_\nu^{(A)}x_A^2 + \lambda_\nu^{(AB)} x_A x_B +\lambda_\nu^{(B)}x_B\,,
\label{A:Hct}
\ee
where the three coefficients (for each bath $\nu$) are now  determined  with the following standard procedure~\cite{ingold02, weiss}.  We consider the minimum of the total Hamiltonian  $H^{(t)}$ (see Eq.~(1) in the main text) with respect to the environment and  the WM coordinates. From the requirement $\frac{\partial H^{(t)}}{\partial X_{k,\nu}} = 0$ 
we obtain
\begin{equation}
\label{eq7}
  X_{k,\nu} =  \sum_{l=A,B} \frac{g_\nu^{(l)}(t)c^{(l)}_{k,\nu}}{m_{k,\nu} \omega_{k,\nu}^2} x_{l}\,.
\end{equation}
This value will be used to determine the minimum of the Hamiltonian with respect to
the WM coordinates $x_l$, which is given by 
\begin{equation}
  \frac{\partial H^{(t)}}{\partial x_l} = \frac{\partial V_{l}(x_l)}{\partial x_{l}} +\frac{\partial H_{{\rm c.t.},\nu}}{\partial x_{l}}
-\!\!\sum_{l=A,B}\sum_{k=1}^{+\infty}g_\nu^{(l)}(t) c_{k,\nu}^{(l)} X_{k,\nu}.
\end{equation}
Now, to avoid renormalization effects we  impose that this minimum corresponds to the minimum of the  bare potential $V_l(x_l)$
\begin{equation}
  \frac{\partial H^{(t)}}{\partial x_l} = \frac{\partial V_{l}(x_l)}{\partial x_{l}}\qquad \text{for } l=A,B\,.
\end{equation}
This constraint results in the following relations 
\be
 \sum_{k=1}^{+\infty} g_\nu^{(l)}(t) c_{k,\nu}^{(l)} X_{k,\nu} = 2\lambda_\nu^{(l)}x_l + \lambda_\nu^{(AB)} x_{\bar l},
\label{eqfinal}
\ee
where we remind  the convention according to which if $l=A$ then $\bar{l} = B$, and \textit{vice versa}.
Inserting now Eq.~\eqref{eq7} into~(\ref{eqfinal}) we get 
\beq
&&\lambda_\nu^{(A)}=\sum_{k=1}^{+\infty} \frac{(g_\nu^{(A)}(t) c_{k,\nu}^{(A)})^2 }{2m_{k,\nu}\omega_{k,\nu}^2};\, \lambda_\nu^{(B)}=\sum_{k=1}^{+\infty} \frac{(g_\nu^{(B)}(t) c_{k,\nu}^{(B)})^2 }{2m_{k,\nu}\omega_{k,\nu}^2}\nonumber\\
&&\lambda_\nu^{(AB)} =\sum_{k=1}^{+\infty} \frac{g_\nu^{(A)}(t)g_\nu^{(B)}(t) c_{k,\nu}^{(A)}c_{k,\nu}^{(B)}}{m_{k,\nu}\omega_{k,\nu}^2}.
\label{A:lambda}
\eeq
The final form  of $H^{(t)}_{{\rm int},\nu}$ is now obtained from the interaction form in Eq.~(\ref{A:Hint}), with  the counter-term given in Eqs.~(\ref{A:Hct}) and~(\ref{A:lambda}). Notice that in this Section, for completeness, we left the explicit  indices $A$ and $B$ in 
$c_{k,\nu}^{(A)}$ and $c_{k,\nu}^{(B)}$. In the following we will consider always $c_{k,\nu}^{(A)}=c_{k,\nu}^{(B)}$.
\section*{Supplementary Note 2: Time dependence evolutions of WM and baths coordinates}
\label{app:eom}
Starting from Eq.~(1) of the main text the equations of motion, in the Heisenberg picture, for the QHO's and reservoir degrees of freedom are, respectively,
\begin{equation}
\begin{cases}
\dot{x}_l \!\!\!\! &=\dfrac{p_l}{m}\,,\\
\dot{p}_l \!\!\!\! &= - m \omega_l^2 x_l - {\displaystyle \sum_{\nu = 1,2}} {\displaystyle \sum_{k=1}^{+\infty}} \bigg[
- g_\nu^{(l)}(t) c_{k,\nu} X_{k,\nu}\\
\!\!\!\! &
+\dfrac{(g_\nu^{(l)}(t) c_{k,\nu})^2}{m_{k,\nu}\omega_{k,\nu}^2} x_{l} 
+ \dfrac{g_\nu^{(l)}(t) g_\nu^{(\bar{l})}(t)  c_{k,\nu}^2  }{ m_{k,\nu}\omega_{k,\nu}^2} x_{\bar{l}} \bigg]\,,
\end{cases}
\label{eq:EoM_DoF_sys}
\end{equation}
and
\begin{equation}
\begin{cases}
\dot{X}_{k,\nu} \!\!\!\! &=\dfrac{P_{k,\nu}}{m_{k,\nu}}\,,\\
\dot{P}_{k,\nu} \!\!\!\! &= - m_{k,\nu} \omega_{k,\nu}^2 X_{k,\nu} + {\displaystyle \sum_{l = A,B}} 
g_\nu^{(l)}(t) c_{k,\nu} x_{l}\,,
\end{cases}
\label{eq:EoM_DoF_res}
\end{equation}
where the overdots denote time derivatives. The solution for the position operator of the $k$-th oscillator of the $\nu$th reservoir is exact and can be written in terms of the WM coordinates
\begin{align}
X_{k,\nu} (t) & = \xi_{k,\nu} (t) + \sum_{l=A,B} \frac{c_{k,\nu}}{m_{k,\nu} \omega_{k,\nu}} \int_{t_0}^{t} \d s \, g_{\nu}^{(l)}(s) x_l(s)\nonumber\\
&\quad \times \sin[\omega_{k,\nu}(t-s)],
\label{eq:X_R_sol_EoM}
\end{align}
where we have introduced
\begin{align}
\xi_{k,\nu}(t) \equiv &  X_{k,\nu} (t_0) \cos [\omega_{k,\nu}(t-t_0)] \nonumber\\
& + \frac{P_{k,\nu}(t_0)}{m_{k,\nu} \omega_{k,\nu}}\sin[\omega_{k,\nu}(t-t_0)],
\label{app:eq:xi_knu_def}
\end{align}
and $t_0\to-\infty$ the initial time. Notice that the corresponding fluctuating noise is
\begin{equation}
\xi_\nu(t)= \sum_{k=1}^{+\infty} c_{k,\nu}\xi_{k,\nu}(t),
\label{A:xi_nu_def1}
\end{equation}
with zero quantum average $\langle \xi_\nu(t)\rangle =0$ and correlator 
\begin{align}
\label{xixi2}
&\langle \xi_\nu(t) {\xi}_{\nu'}(t') \rangle = \delta_{\nu,\nu'}\int_0^\infty \frac{\d\omega}{\pi} {\cal J}_\nu (\omega)  \nonumber\\
& \times \left[ \coth\left(\frac{\omega}{2 T_\nu}\right) \cos[\omega (t-t')] - i \sin[\omega (t-t')]\right].
\end{align}
To determine the time behaviour of $x_l(t)$ we remind that we have assumed the coupling to the reservoir $\nu=1$ much weaker  \red{than} the one with $\nu=2$. In this spirit, we  will treat the interaction term $H^{(t)}_{\rm int, \nu=1}$ perturbatively. We can then split the total Hamiltonian $H^{(t)}$ in Eq.~(1) of the main text as $H^{(t)}=H^{(0)}+
H^{(t)}_{\rm int, \nu=1}$, where  $H^{(0)}$ is the  "unperturbed" part, which contains all the contributions of $H^{(t)}$ apart $H^{(t)}_{\rm int, \nu=1}$ which is indeed the perturbation.
Within this procedure the time dependence of the position operator of the $l$-th QHO is 
\begin{equation}
x_l(t)= x_l^{(0)}(t) + \Delta x_l(t).
\label{app:x_qho_pert} 
\end{equation}
Here, the first term is the leading one and  it evolves under the influence of the static $\nu=2$ reservoir
\begin{equation}
x_l^{(0)}(t)  = e^{i H^{(0)}(t-t_0)} x_{l}^{(0)} e^{-i H^{(0)}(t-t_0)}~.
\label{eq:xl0_qho}
\end{equation}
On the other hand, the second term is the perturbative correction evaluated  at lowest order in $H^{(t)}_{\rm int, \nu=1}$
\begin{align}
\Delta x_l(t) &=-i \int_{t_0}^{t} \d s\sum_{l'=A,B} g_1^{(l')}(s) \xi_1(s)\big[x_{l'}^{(0)}(s),x_l^{(0)}(t)\big]~.
\label{app:delta_xl_qho}
\end{align}
From the above expressions it is clear that the time dependence of $x_{A,B}(t)$ is directly obtained from the time behaviour of the leading terms ${x}^{(0)}_{A,B}(t)$. These latter operators obey the coupled Langevin equations in the presence of the coupling with the static bath $\nu=2$. We have ($l=A,B$; $\bar l=B,A$)
\begin{align}
\label{app_langevin2}
\ddot{x}^{(0)}_l(t) &+ \omega_l^2 x^{(0)}_l(t) + \int_{-\infty}^t \d s~\Big[\gamma_2(t-s)\dot{x}^{(0)}_l(s) \nonumber\\
&+ \gamma_2(t-s)\dot{x}^{(0)}_{\bar l}(s)\Big]=\frac{\xi_2(t)}{m}\,,
\end{align}
with damping kernel
\begin{equation}
\gamma_\nu(t)=\frac{1}{m}\sum_{k=1}^{+\infty} \frac{c_{k,\nu}^2}{m_{k,\nu}\omega_{k,\nu}^2}\cos(\omega_{k,\nu}t)~,
\label{eq:gamma}
\end{equation}
and noise $\xi_2(t)$ given in Eq.~(\ref{A:xi_nu_def1}). The formal solution, in the long time limit, can be written as a convolution
\be 
\label{convolution}
x^{(0)}_l(t)=\sum_{l'=A,B}\frac{1}{m}\int_{-\infty}^{+\infty}\d s \chi_2^{(l, l')}(t-s)\xi_2(s)~,
\ee
where we have introduced the retarded response function (see also \red{App. B})
\be
\label{appresponsefunction}
\chi^{(l l')}_2(t)\equiv i m \theta(t) \langle [x^{(0)}_l(t), x_{l'}^{(0)}(0)] \rangle ~.
\ee
Notice that here the quantum averages $\langle \dots\rangle$ are performed only with respect to the static $\nu=2$ reservoir contribution and can be taken as thermal averages since we are interested in the long time behaviour.
\section*{Supplementary Note 3: Perturbative expressions of thermodynamic quantities}
\label{app:perturbation}
In this Section we derive closed expressions for the average power and heat currents under the assumption that the (time-dependent) couplings with the $\nu=1$ reservoir are weak. We start by considering the power expression \red{in Eq.~(A1) in App. A}. Inserting the solution $X_{k,1}(t)$ of Eq.~\eqref{eq:X_R_sol_EoM}, and after an integration by parts (exploiting also that $g_1(t_0) = 0$ because the coupling is switched on at $t_0^+$), we obtain
\begin{align} 
& {P}(t) =  -\sum_{l=A,B}
\dot{g}_1^{(l)}(t) x_l(t)  \bigg\lbrace\xi_{1}(t) \nonumber\\
& -\sum_{l' = A,B} m \int_{t_0}^{t} \d s\, \gamma_1(t-s) \frac{\d}{\d s} [ g_1^{(l')}(s) x_{l'}(s) ] \bigg\rbrace\,,
\label{eq:power_op_exact}
\end{align}
with the damping function defined in Eq.~\eqref{eq:gamma}.
Now, using Eq.~\eqref{app:x_qho_pert}, we write $P(t)$ at the lowest perturbative order 
\begin{align}
&{P}(t) = -\!\! \sum_{l=A,B} \bigg\lbrace 
\! \dot{g}_1^{(l)}(t) [x_l^{(0)}(t) + \Delta x_l(t)] \xi_{1}(t)\nonumber\\
&- \!\dot{g}_1^{(l)}(t) x_l^{(0)}(t) \!\!\!\! \sum_{l' = A,B} m\int_{t_0}^{t} \!\!\! \d s\, \gamma_1(t-s) \frac{\d}{\d s} [ g_1^{(l')}(s) x_{l'}^{(0)}(s) ] \! \bigg\rbrace .
\label{eq:pwr_pert_op}
\end{align}
In order to perform the quantum average $\langle P(t)\rangle$ we first notice that  $\langle x_l^{(0)}(t) \dot{\xi}_1(t') \rangle = 0$, since  at this perturbative order $ x_l^{(0)}(t)$ and $\dot{\xi}_1(t)$  are completely decoupled.
In addition, it is useful  to introduce in Eq.~(\ref{eq:power_op_exact}) the WM correlators
\be
C^{(l,l')}(t,s)\equiv C^{(l, l')}(t-s)=\langle x_l^{(0)}(t)x_{l'}^{(0)}(s)\rangle~,
\label{c_correlator}
\ee
with combinations
\be
\label{cpm}
C_\pm^{(l,l')}(\tau)= C^{(l,l')}(\tau) \pm C^{(l',l)}(-\tau)~,
\ee
and  to change variable $\tau=t-s$. After these steps it is then straighforward to  perform the average $\langle P(t)\rangle$  also over the period. Defining  the function
\be
\label{g_grande1}
G^{(l,l')}(\tau)=\int_{{0}}^{{\cal T}}\frac{\d t}{{\cal T}}\dot g_1^{(l)}(t) g_1^{(l')}(t-\tau)\,,
\ee
we arrive at 
\beq
\label{app_p_pert1}
P=&&-i\!\sum_{l,l'=A,B}\int_0^{+\infty}\!\!\!\!\!\d\tau\,G^{(l,l')}(\tau)\Big\lbrace {\cal L}_{1,{\rm s}}(\tau)C_-^{(l,l')}(\tau) \nonumber\\
&&+{\cal L}_{1,{\rm a}}C_+^{(l,l')}(\tau) \Big\rbrace~.
\eeq
In the above expression, we  have   ${\cal L}_1(\tau) \equiv {\cal L}_{1,{\rm s}}(\tau) + {\cal L}_{1,{\rm a}}(\tau)$ with symmetric and anti-symmetric parts
\beq
\label{lsla}
{\cal L}_{1,{\rm s}}(t)&=&\int_0^{+\infty}\!\!\frac{\d\omega}{\pi} 
{\cal J}_1(\omega)\coth(\frac{\omega}{2T_1})\cos(\omega t),\nonumber\\
{\cal L}_{1,{\rm a}} (t)&=&-i \int_0^{+\infty}\!\!\frac{\mathrm{d}\omega}{\pi} 
{\cal J}_1(\omega)\sin(\omega t)~.
\eeq
Notice that ${\cal L}_1(\tau)$ corresponds to the noise correlator $\langle \xi_1(\tau)\xi_1(0)\rangle$ defined in Eq.~(\ref{xixi2})
and depends on the bath spectral density ${\cal J}_1(\omega)$. For the sake of completeness we quote also the identity
$m\dot{\gamma}_1(t)=-2i {\cal L}_{1,{\rm a}}(t)$. To obtain now the final expression for the power we use the  Fourier components of  the functions ${\cal L}_{1,{\rm {s,a}}}(\tau)$ and  $C_\pm^{(l,l')}(\tau)$. They are 
\beq
{\cal L}_{1,{\rm s}} (\omega)&=&{\cal J}_1(\omega)\coth\left(\frac{\omega}{2T_1}\right);\quad
{\cal L}_{1,{\rm a}}(\omega)={\cal J}_1(\omega)\nonumber\\
C_+^{(l,l')}(\omega) &=& \frac{2}{m}\coth\left(\frac{\omega}{2T_2}\right){\chi_2^{(l,l')}}''(\omega)~,\nonumber \\
C_-^{(l,l')}(\omega) &=& \frac{2}{m}{\chi_2^{(l,l')}}''(\omega)~.
\eeq
For $C_\pm^{(l,l')}(\omega)$ we used the  fluctuation dissipation relation that links these correlators  to the imaginary part of the response function $\chi_2^{(l,l')}(\omega)$. Concerning $G^{(l,l')}(\tau)$ in Eq.~(\ref{g_grande1}) we use the Fourier decomposition 
\be
\label{eq:decomposition}
g_1^{(l)}(t)=\sum_{n=-\infty}^{+\infty} g_n^{(l)}e^{-i n\Omega t}
\ee
to find
\be
\label{g_grande2}
G^{(l,l')}(\tau)=-i\sum_{n=-\infty}^{+\infty} n\Omega g_n^{(l)}g_{-n}^{(l')}e^{-in \Omega\tau}~.
\ee
Exploiting these Fourier representations in Eq.~\eqref{app_p_pert1}, one eventually arrives at the final expression for the average power 
\be
P \! = \! - \!\!\!\! \sum_{n=-\infty}^{+\infty} \!\!\! n\Omega \!\! \int_{-\infty}^{+\infty} \!\!\! \frac{\d\omega}{2\pi m} {\cal J}_1(\omega+n\Omega) N(\omega,n\Omega) {\bf g}^\dag_n\cdot {\bm \chi}_2''(\omega)\cdot {\bf g}_n\,, 
\ee
where ${\bf g}_n=(g_n^{(A)}, g_n^{(B)})$ is a two-element vector and ${\bf g}_n^\dag$ its adjoint. We have also used the compact matrix notation 
${\bm \chi}_2''(\omega)$, which is a two-by-two matrix whose entries are the imaginary parts of the response functions ${\chi_2^{(l,l')}}(\omega)$ taken from Eq.~\eqref{appresponsefunction}, and introduced
\begin{equation}
\label{app_n_function}
N(\omega, \Omega) = \coth\left(\frac{\omega + \Omega}{2T_1}\right)-\coth\left(\frac{\omega}{2T_2}\right)\,.
\end{equation}

Concerning the average heat currents $J_\nu$, similar steps as detailed above can be followed. Here, we briefly report few of them  to arrive to the final closed expressions.
 Considering 	\red{Eq.~(A2) of App.A} and expanding $x_l(t)$ as in Eq.~\eqref{app:x_qho_pert}, we have 
\begin{align}
&J_1(t) = - \!\!\sum_{l = A,B} g_{1}^{(l)}(t) \biggl[ x_l^{(0)}(t) \dot{\xi}_1^{(l)}(t) + \Delta x_l(t) \dot{\xi}_1^{(l)}(t) \nonumber\\
& -m  \!\!\sum_{l' = A,B} \int_{t_0}^{t} \d s\, x_l^{(0)}(t) x_{l'}^{(0)}(s) g_{1}^{(l')}(s) \ddot{\gamma}_{1}(t-s)\biggr].
\end{align}
After taking the quantum average, the lowest order perturbative expression reads
\begin{align}
&\langle J_1(t)\rangle
 =\sum_{l,l'=A,B}  \int_{t_0}^{t} \d s\, g_{1}^{(l)}(t) g_{1}^{(l')}(s)\biggl\lbrace m\ddot{\gamma}_{1}(t-s) \nonumber\\
& \times C^{(l,l')}(t-s) -i \langle \xi_1(s) \dot{\xi}_1(t) \rangle  C_-^{(l,l')}(t-s) \biggr\rbrace ~.
\end{align}
As before, the average over one period of the drive is performed by introducing the correlators in Eq.~\eqref{cpm}, exploiting the Fourier decomposition in 
Eq.~\eqref{eq:decomposition} and using Eqs. \eqref{g_grande1}-\eqref{g_grande2}. 
After some straightforward steps one arrives at
\begin{align}
J_1 \! = \!\sum_{n=-\infty}^{+\infty}  \int_{-\infty}^{+\infty} & \frac{\d\omega}{2\pi m} (\omega + n\Omega) {\cal J}_1(\omega+n\Omega) N(\omega,n\Omega) \nonumber\\
&\times {\bf g}^\dag_n\cdot{\bm \chi}''_2 (\omega)\cdot{\bf g}_n~,
\end{align}
and similarly for the other heat current
\be
J_2 \! = \! - \!\!\!\! \sum_{n=-\infty}^{+\infty} \! \int_{-\infty}^{+\infty} \!\!\! \frac{\d\omega}{2\pi m} \omega  {\cal J}_1(\omega+n\Omega) N(\omega,n\Omega) {\bf g}^\dag_n\cdot{\bm \chi}''_2(\omega)\cdot{\bf g}_n~.
\ee

 \section*{Supplementary Note 4: General properties of the Pareto front}
\label{app:pareto}
In this Section we derive general properties of the Pareto-front and of the corresponding optimal driving. In particular, we prove that: (i) Pareto optimal drives for $(\eta,-P)$ are also optimal for $(\sigma,-P)$; (ii) the $(\sigma,-P)$ Pareto front is weakly convex; (iii) the strictly convex (linear) part of the Pareto front can be described by real coefficients $g_n^{(l)}$ involving at most two (four) frequencies, \ie they are zero except for two (four) values of $n$; (iv) the time-average of Pareto optimal drives is zero, \ie $g_0^{(l)}=0$.
	
Since the time-dependent controls $g_1^{(l)}(t)$ are real, their Fourier coefficients satisfy the property $g_{-n}^{(l)}  = (g_{n}^{(l)})^*$. Using this, and the fact that $\chi_2^{(A B)}(\omega) = \chi_2^{(B A)}(\omega)$, we can rewrite the power and entropy production rate, as
\begin{align}
	P(\{ {\bf g}_n \}) &= \sum_{n=1}^{+\infty} {\bf g}_n^\dagger \cdot I_n^{(P)} \cdot {\bf g}_n , 
	\label{eq:app_p_n} \\
	\sigma(\{ {\bf g}_n \}) &= \sigma_0 + \sum_{n=1}^{+\infty} {\bf g}_n^\dagger \cdot I_n^{(\sigma)} \cdot {\bf g}_n,
	\label{eq:app_sigma_n}
\end{align}
where the sums extend only over positive frequencies and $I_n^{(P)}$ and $I_n^{(\sigma)}$ are 2x2 real and symmetric matrices given by 
\begin{multline}
	 \left[I^{(P)}_n\right]_{ll^\prime} = -(n\Omega)\,  
	 \int_{-\infty}^{+\infty} \frac{\d\omega}{2\pi m} \Big[\mathcal{J}_1(\omega+n\Omega)N(\omega,n\Omega) \\
	 - \mathcal{J}_1(\omega-n\Omega)N(\omega,-n\Omega)\Big] { \chi_2^{(l l^\prime)}}''(\omega) ,
\end{multline}
\begin{multline}
	 \left[I^{(\sigma)}_n\right]_{ll^\prime} = -\frac{1}{T_1}\left[I^{(P)}_n\right]_{ll^\prime} +  \left(\frac{1}{T_2}-\frac{1}{T_1}  \right)  \int_{-\infty}^{+\infty} \frac{\d\omega}{2\pi m}\,\omega \\
	 \times \Big[\mathcal{J}_1(\omega+n\Omega)N(\omega,n\Omega)  + \mathcal{J}_1(\omega-n\Omega)N(\omega,-n\Omega)\Big] \\
	 \times  {\chi_2^{(ll^\prime)}}''(\omega) ,
\end{multline}
for $l,l^\prime = \{A,B\}$.
In Eq.~(\ref{eq:app_sigma_n}) $\sigma_0 = (1/2) {\bf g}_0^\dagger \cdot I_0^{(\sigma)} \cdot {\bf g}_0$ is the entropy production rate given by the zero-frequency driving, \ie the entropy production rate when the coupling is constant in time. The second law of thermodynamics prescribes $\sigma_0\geq 0$. Since the zero frequency component ${\bf g}_0$ does not contribute to the power and it only has a detrimental effect on the entropy production rate, all Pareto-optimal drivings will have ${\bf g}_0 = 0$. From now on we thus set ${\bf g}_0 = 0$.

We now define the $2N_\text{max}$-component vector
\begin{equation}
	{\bf  g} =
	\begin{pmatrix}
		g_{1}^{(A)}, & g_{1}^{(B)},  & g_{2}^{(A)}, & g_{2}^{(B)}, & \dots, & g_{N_\text{max}}^{(A)},  & g_{N_\text{max}}^{(B)}
	\end{pmatrix}^t,
\end{equation}
where $N_\text{max}$ is a maximum cut-off value (not to be confused with $\omega_c$) introduced for numerical purposes. With this definition, ${\bf g}$ is given by the concatenation of ${\bf g}_n$ for all $n=1,2,\dots, N_\text{max}$. 
Let us further introduce two $(2N_\text{max})$x$(2N_\text{max})$ matrices, $I^{(P)}$ and $I^{(\sigma)}$, such that Eqs.~(\ref{eq:app_p_n}) and~(\ref{eq:app_sigma_n}) can be written as
\be
\begin{aligned}
	P({\bf g}) &= {\bf g}^\dagger \cdot I^{(P)} \cdot {\bf g}, &
	\sigma({\bf g}) &= {\bf g}^\dagger \cdot I^{(\sigma)} \cdot {\bf g}.
\end{aligned}
\label{eq:p_sigma_def}
\ee
Importantly, $I^{(P)}$ and $I^{(\sigma)}$ are block-diagonal matrices with 2x2 blocks along the diagonal given by, respectively, $I^{(P)}_1$, $I^{(P)}_2$, \dots $I^{(P)}_{N_\text{max}}$ and  $I^{(\sigma)}_1$, $I^{(\sigma)}_2$, \dots $I^{(\sigma)}_{N_\text{max}}$. Therefore, they have the property that they only mix components with the same frequency index $n$.

Our goal is to characterize the $(\sigma,-P)$ and the $(\eta,-P)$ Pareto fronts assuming that $I^{(P)}$ and $I^{(\sigma)}$ are fixed matrices. The optimization is performed with respect to the vector ${\bf g}$ with $2N_\text{max}$ components. We assume that $\Omega$ is a sufficiently small frequency such that $n\Omega$ is approximately continuous. We then consider a large value of $N_\text{max}$ such that $N_\text{max}\Omega=0.5\omega_A$. This corresponds to optimizing the performance of the quantum thermal machine with respect to an arbitrary driving. The optimization is carried out imposing the two constraints on the Fourier coefficients detailed in the main text.
These constraints can be conveniently written in terms of $\bf{g}_n$ as
\be
\begin{aligned}
	\sum_{n=1}^{+\infty}{\bf g}_n^\dagger \cdot \Pi^{(A)}_n \cdot {\bf g}_n &= |g^{(A)}|^2, &
	\sum_{n=1}^{+\infty}{\bf g}_n^\dagger \cdot \Pi^{(B)}_n \cdot {\bf g}_n &= |g^{(B)}|^2,
\end{aligned}
\ee
where 
\begin{align}
	\Pi^{(A)}_n &= 
	\begin{pmatrix}
		1 & 0 \\ 0 & 0
	\end{pmatrix},
	&
	\Pi^{(B)}_n &= 
	\begin{pmatrix}
		0 & 0 \\ 0 & 1
	\end{pmatrix},
\end{align}
project on the corresponding subspace relative to QHO $A$ or $B$. 
We can write the constraints in terms of ${\bf g}$ as
\be
\begin{aligned}
	{\bf g}^\dagger \cdot \Pi^{(A)} \cdot {\bf g} &= |g^{(A)}|^2, &
	{\bf g}^\dagger \cdot \Pi^{(B)} \cdot {\bf g} &= |g^{(B)}|^2, 
\end{aligned}
\label{eq:app_constraints}
\ee
where 
the projector $\Pi^{(A)}$ ($\Pi^{(B)}$) is a $(2N_\text{max})$x$(2N_\text{max})$ diagonal matrix with $\Pi^{(A)}_n$ ($\Pi^{(B)}_n$) along its diagonal. 
We notice that if there was a single constraint on the square norm of ${\bf g}$, the optimization of the power and entropy production rate would be straightforward, because they are quadratic forms, and their maximum (minimum) would be given by the largest (smallest) eigenvalue. However, as we now show, the presence of $2$ constraints and the search of the full Pareto front introduces a richer set of solutions.
\subsection*{Pareto optimal drives for $(\eta,-P)$ are also Pareto optimal for $(\sigma,-P)$ (but not vice-versa)}
We now show that if a drive is optimal for $(\eta,-P)$, then it is also optimal for $(\sigma,-P)$. This implies that if we find the full $(\sigma,-P)$ Pareto front, and we transform these points to $(\eta,-P)$, we will have a set of points that necessarily contains the full $(\eta,-P)$ Pareto front. Indeed, in Figures~\red{4 and ~5} of the main text the plots in the right column are found by transforming all points in the left column, and selecting only those that are Pareto-optimal.

The transformation between the two Pareto fronts is given by the identity
\be
	\eta(\sigma,P) = \eta_\text{C}\left[1- \frac{\sigma T_2}{P} \right]^{-1}.
	\label{eq:eta_p_sigma}
\ee

We now use Eq.~(\ref{eq:eta_p_sigma}) to prove our statement by contradiction. Let ${\bf g}$ be a driving that is Pareto optimal for power and efficiency, \ie such that $[\eta(\sigma({\bf g}),P({\bf g})),-P({\bf g})]$ is on the Pareto front. If, by contradiction, $[\sigma({\bf g}),-P({\bf g})]$ was not also on the Pareto front for power and entropy production, then there would be a driving $\bar{\bf g}$ such that $[\sigma(\bar{\bf g}),-P(\bar{\bf g})]$ is strictly better than $[\sigma({\bf g}),-P({\bf g})]$. This means that either
\begin{align}
	P(\bar{\bf g}) &< P({\bf g}) & \sigma(\bar{\bf g}) &\leq \sigma({\bf g}) ,
	\label{eq:app_p1}
\end{align}
or
\begin{align}
	P(\bar{\bf g}) &\leq P({\bf g}) & \sigma(\bar{\bf g}) &< \sigma({\bf g}).
	\label{eq:app_p2}
\end{align}

If Eq.~(\ref{eq:app_p1}) is true, using the monotonicity of $\eta(P,\sigma)$ with respect to $P$ and $\sigma$, and using that $\sigma >0$ and $P<0$ in the heat engine regime, we have that
\begin{equation}
	\eta(\sigma(\bar{\bf g}),P(\bar{\bf g})) \geq \eta(\sigma({\bf g}),P(\bar{\bf g}))  > \eta(\sigma({\bf g}),P({\bf g})).
	\label{eq:app_eta1}
\end{equation}
Eqs.~(\ref{eq:app_p1}) and~(\ref{eq:app_eta1}) imply that the driving ${\bf g}$ has a strictly worse extracted power and efficiency with respect to $\bar{\bf g}$, which would be a contradiction to our hypothesis. Therefore Eq.~(\ref{eq:app_p1}) cannot be true.
A similar argument can be used if  Eq.~(\ref{eq:app_p2}) is true, thus concluding our proof by contradiction.

\subsection*{The outer strictly convex part of the $(\sigma,-P)$ Pareto front can be described by real coefficient $g_n^{(l)}$ involving at most two frequencies}
We can determine the outer strictly convex part of the $(\sigma,-P)$ Pareto front maximizing the figure of merit 
\begin{equation}
	F_c({\bf g}) = -c P({\bf g}) - (1-c) \sigma({\bf g}) = {\bf g}^\dagger \cdot I^{(F)}_c \cdot {\bf g}
\end{equation}
for all values of $c\in [0,1]$~\cite{seoane2016}, where 
\be
	I^{(F)}_c = -c I^{(P)} - (1-c)I^{(\sigma)}
\ee
is a block-diagonal, real and symmetric matrix.
Graphically, the optimization of $F_c({\bf g})$ corresponds to finding the outermost points of the Pareto front that are tangent to the vector $(c,1- c)$ in the $(\sigma,-P)$ plane.
To enforce the constraints in Eq.~(\ref{eq:app_constraints}), we introduce the following Lagrangian
\begin{multline}
	L_c({\bf g}, \lambda^{(A)}, \lambda^{(B)}) = \frac{1}{2}F_c({\bf g})  \\
	-\frac{1}{2} \sum_{l=A,B} \lambda^{(l)}\left[ {\bf g}^\dagger \cdot \Pi^{(l)} \cdot {\bf g} - |g^{(l)}|^2 \right].   
\end{multline}
The derivative of $L_c({\bf g}, \lambda^{(A)}, \lambda^{(B)}) $ with respect to $\lambda^{(A)}$ and $\lambda^{(B)}$ imposes the constraints in Eq.~(\ref{eq:app_constraints}). Since the driving ${\bf g}$ is, in principle, a complex, we express it as ${\bf g} = {\bf a} + i{\bf b}$. The optimality conditions are then obtained deriving the Lagrangian with respect to the real parameters ${\bf a}$ and $\bf{ b}$. 
Using that $I_c^{(F)}$, $\Pi^{(A)}$ and $\Pi^{(B)}$ are real and symmetric matrices, and that $\Pi^{(A)}+\Pi^{(B)}=\text{Id}$, it can be shown that the optimality condition for the complex vector ${\bf g}$ can be expressed as
\begin{equation}
	\left[ I^{(F)}_c + \lambda_d \left(\Pi^{(A)} -\Pi^{(B)}  \right) \right]{\bf g} = \lambda_s {\bf g},
	\label{eq:optim_pareto}
\end{equation}
where, for convenience, we introduced $\lambda_d = (\lambda^{(B)}-\lambda^{(A)})/2$ and $\lambda_s =  (\lambda^{(B)}+\lambda^{(A)})/2$.

Let $(\bar{\bf g}, \bar{\lambda}_s, \bar{\lambda}_d)$ be an optimal driving. It must satisfy Eq.~(\ref{eq:optim_pareto}) and the constraints in Eq.~(\ref{eq:app_constraints}).  
From Eq.~(\ref{eq:optim_pareto}), we see that $\bar{\bf g}$ is an eigenvector of
\begin{equation}
 	M = I^{(F)}_c + \bar{\lambda}_d \left(\Pi^{(A)} -\Pi^{(B)}  \right)
 	\label{eq:m_def}
\end{equation}
with eigenvalue $\bar{\lambda}_s$.
Since $M$ is real and symmetric, there will be an orthonormal basis of \textit{real} vectors $\{{\bf  e_i }\}_i$ that generate the eigenspace of $M$ relative to the eigenvalue $\bar{\lambda}_s$. Furthermore, since $M$ is block-diagonal with 2x2 matrices along the diagonal, we can assume that each eigenstate ${\bf e_i} $ has at most two non-null components corresponding to a single frequency index $n$.

 In general, $\bar{\bf g}$ will be a \textit{complex} linear combination of  $\{{\bf e_i} \}_i$, i.e.
 \begin{equation}
 	\bar{\bf g} = \sum_i \bar{\zeta}_i {\bf e_i}.
 	\label{eq:app_g_step1}
 \end{equation}
 As we prove below,  there is a linear combination $\{ \zeta_i \}_i$ of the $\{{\bf e_i}\}_i$ vectors that (i) is real, (ii)
satisfies the constraints, (iii) has the same value of the figure of merit $F_c(\bar{{\bf g}})$, and (iv) has at most $2$ non-null components.

 Having the same figure of merit means that the corresponding power and entropy production lie on the tangent of the Pareto front in $(\sigma(\bar{\bf g}),-P(\bar{\bf g}))$. However, since we are characterizing the outer strictly convex part of the Pareto front, the only point that can lie along the tangent in $(\sigma(\bar{\bf g}),-P(\bar{\bf g}))$, satisfying the constraints, is the point $(\sigma(\bar{\bf g}),-P(\bar{\bf g}))$ itself. Therefore, we can equivalently pick the real linear combination of real vectors ${\bf g} = \sum_i  \zeta_i  {\bf e_i}$ to represent the $(\sigma(\bar{\bf g}),-P(\bar{\bf g}))$ point on the Pareto front, proving the statement of this Subsection.

We now prove the existence of $\{\zeta_i\}_i$.
First we notice that if ${\bf g} = \sum_i \zeta_i {\bf e_i}$ satisfies the constraints in Eq.~(\ref{eq:app_constraints}), we have that 
 \begin{equation}
	F_c({\bf g}) = \left( \bar{\lambda}_s-\bar{\lambda}_d \right)|g^{(A)}|^2 + \left( \bar{\lambda}_s+\bar{\lambda}_d \right)|g^{(B)}|^2.
	\label{eq:fc_s1}
\end{equation}
This can be derived multiplying Eq.~(\ref{eq:m_def}) by ${\bf g}^\dagger$ and ${\bf g}$ respectively from the left and right. Notably, in Eq.~(\ref{eq:fc_s1}) $F_c({\bf g})$ does not depend on the specific linear combination. Therefore, any linear combination  ${\bf g} = \sum_i \zeta_i {\bf e_i}$ satisfying the constraints in Eq.~(\ref{eq:app_constraints}) will have the same value of the figure of merit of $F(\bar{\bf g})$.

We now show that there is always a real linear combination $\{\zeta_i\}_i$, satisfying the constraints in Eq.~(\ref{eq:app_constraints}), with at most two non-null coefficients, concluding the proof of this Subsection. For $O =\Pi^{(A)}, \Pi^{(B)}$, let us consider the identity
\begin{equation}
	\left( {\bf g} \right)^\dagger \cdot O \cdot  {\bf g} = \sum_i |\zeta_i|^2 {\bf e_i}^t\cdot O\cdot {\bf e_i} + \sum_{i\neq j} \zeta_i^*z_j {\bf e_i}^t \cdot O \cdot {\bf e_j}.
	\label{eq:app_o_step1}
\end{equation}
We impose that the coefficients $\{\zeta_i\}_i$ satisfy the constraints in Eq.~(\ref{eq:app_constraints}). Using that ${\bf e_i}^t \cdot O \cdot {\bf e_j}=0$ (proven below), this yields, for $l=A,B$,
\begin{equation}
	\sum_i P_{li} |\zeta_i|^2  = |g^{(l)}|^2,
	\label{eq:constraints}
\end{equation}
where $P_{li} = {\bf e_i}^t \cdot \Pi^{(l)} \cdot {\bf e_i}$.
Equation~(\ref{eq:constraints}) is a linear problem for the coefficients $|\zeta_i|^2$. Since Eq.~(\ref{eq:constraints}) only depends on the square modulus of $\zeta_i$, we can equivalently assume that $\zeta_i$ is real. Since $P_{li}$ is a $2$x$m$ matrix, where $m$ is the number of elements of $\{ \bf e_i \}_i$, it has at most rank $2$. Combined with the fact that a non-null solution exist, \ie $\bar{\bf g}=\sum_i \bar{z}_i {\bf e_i}$, it can be shown that it is always possible to find a solution with at most two non-null coefficients. 

To conclude the proof, we show that ${\bf e_i}^t \cdot O \cdot {\bf e_j}=0$ for all $i\neq j$. We distinguish two cases. If ${\bf e_i}$ and ${\bf e_j}$ correspond to different frequency indices $n$, then  ${\bf e_i}^t \cdot O \cdot {\bf e_j}$ is trivially null, because $\Pi^{(A)}, \Pi^{(B)}$ are diagonal. If they correspond to the same frequency index $n$, let us denote with  ${\bf e^{(n)}_{i}}$ and ${\bf e^{(n)}_{j}}$ the two dimensional vector containing the non-null components of  ${\bf e_i}$ and ${\bf e_j}$, respectively. ${\bf e^{(n)}_{i}}$ and ${\bf e^{(n)}_{j}}$ are orthonormal eigenvectors of the corresponding 2x2 matrix 
\begin{equation}
	M_n =  \left[I^{(F)}_c\right]_n + \bar{\lambda}_d \left(\Pi^{(A)}_n -\Pi^{(B)}_n  \right),
	\label{eq:app_mn_step1}
\end{equation}
relative to the \textit{same} eigenvalue $\bar{\lambda}_s$. Here
 \begin{equation}
 	\left[I^{(F)}_c\right]_n = -c I^{(P)}_n -(1-c)I^{(\sigma)}_n.
 \end{equation}
 Since $M_n$ is a 2x2 matrix with two identical eigenvalues $\bar{\lambda}_s$, it can be shown that $M_n = \bar{\lambda}_s \text{Id}$. Since $M_n$ is diagonal, without loss of generality we can choose ${\bf e_i}^{(n)}= (1,0)$ and ${\bf e_j}^{(n)}=(0,1)$. 
With this choice, ${\bf e_i}^t \cdot O \cdot {\bf e_j}$ is zero because $\Pi^{(A)}_n$ and $\Pi^{(B)}_n$ are diagonal matrices.

\subsection*{The $(\sigma,-P)$ Pareto front is weakly convex and generated by real coefficients $g_n^{(l)}$ involving at most two (four) frequencies where it is strictly convex (linear)}
Let us consider the outer strictly convex part of the Pareto front identified as in the previous Subsection. Below we prove that, mixing at most $4$ frequencies with real coefficients $g_n^{(l)}$, we can find solutions generating the entire convex hull of the outer strictly convex part. The entire Pareto front is then given by the outer border of the convex hull, which in turns is composed of a strictly convex part (mixing at most two frequencies with real coefficients -- as proven in the previous Subsection), and segments (mixing at most four frequencies with real coefficients). Indeed, if by contradiction there was a strictly better solution outside of the convex hull, which is a weakly convex curve, there would be a value of $c$ such that the maximum of $F_c({\bf g})$ in Eq.~(\ref{eq:fc_s1}) coincides with such a solution. By hypothesis, this point cannot be part of the convex hull, but all points maximizing $F_c({\bf g})$ generate the convex hull, and are thus part of it. This concludes the proof.

We now show that, given two drivings ${\bf g}^{(1)}$ and ${\bf g}^{(2)}$ belonging to the outer strictly convex part of the Pareto front, we can find drivings with real coefficients mixing at most four frequencies that generate the entire segment between $( \sigma({\bf g}^{(1)}),-P({\bf g}^{(1)}))$ and $(\sigma({\bf g}^{(2)}),-P({\bf g}^{(2)}))$.

Using the results of the previous Subsection, ${\bf g}^{(1)}$ and ${\bf g}^{(2)}$ satisfy the constraints in Eq.~(\ref{eq:app_constraints}) and can be expressed as real combination of two frequencies, \ie
\be
\begin{aligned}
	{\bf g}^{(1)} &= x_{11} {\bf g}^{(11)} + x_{12} {\bf g}^{(12)},\\
	{\bf g}^{(2)} &= x_{21} {\bf g}^{(21)} + x_{22} {\bf g}^{(22)},
\end{aligned}
\label{eq:app_s1}
\ee
where ${\bf g}^{(ij)}$, for $i,j=1,2$, are real vectors with a single frequency and $x_{ij}$ are real coefficients.

Let us consider the drive
\begin{equation}
	\bar{\bf g}(a) = \sqrt{a} {\bf g}^{(1)} + \sqrt{1-a} {\bf g}^{(2)}
\end{equation}
which, by construction, has real coefficients and mixes at most $4$ frequencies.
For $O=I^{(P)}, I^{(\sigma)}, \Pi^{(A)}, \Pi^{(B)}$ let us compute
\begin{multline}
	\bar{\bf g}(a)^\dagger \cdot O \cdot \bar{\bf g}(a) = a {\bf g}^{(1)\dagger} \cdot O \cdot {\bf g}^{(1)} 
	+ (1-a){\bf g}^{(2)\dagger} \cdot O \cdot {\bf g}^{(2)}  \\
	+ 2\sqrt{a}\sqrt{a-1} {\bf g}^{(1)\dagger} \cdot O \cdot {\bf g}^{(2)}.
\end{multline}
If ${\bf g}^{(1)\dagger} \cdot O \cdot {\bf g}^{(2)}=0$, $\bar{\bf g}(a)$ satisfies both constraints in Eq.~(\ref{eq:app_constraints}), and its power and entropy production rate linearly interpolate between $(\sigma({\bf g}^{(1)}),-P({\bf g}^{(1)}))$ and $(\sigma({\bf g}^{(2)}),-P({\bf g}^{(2)}))$ for $a\in [0,1]$. This would conclude the proof.
From Eq.~(\ref{eq:app_s1}), we see that ${\bf g}^{(1)\dagger} \cdot O \cdot {\bf g}^{(2)}=0$ if each term
\begin{equation}
	{\bf g}^{(1i)\dagger} \cdot O \cdot {\bf g}^{(2j)} = 0,
	\label{eq:s2}
\end{equation}
for $i,j=1,2$. If ${\bf g}^{(1i)}$ and ${\bf g}^{(2j)}$ belong to different frequencies, then Eq.~(\ref{eq:s2}) is true because $O=I^{(P)}, I^{(\sigma)}, \Pi^{(A)}, \Pi^{(B)}$ are block diagonal. If instead they belong to the same frequency, we can change one of the two frequencies by an infinitesimal amount. This will have an arbitrarily small effect on the performance and on the constraints, but guarantees that Eq.~(\ref{eq:s2}) holds.

\section*{Supplementary Note 5: Numerical approach to find the Pareto front}
\label{app:numeric}
Here we provide some details on the numerical approach used to compute the Pareto fronts shown in Figures~\red{4} and \red{5}  of the main text and we show some representative examples of the distribution of $g_n^{(l)}$ found with this optimization method.

As described in the main text, we must solve numerically \red{the optimization problem in Eq.~(21) subject to three constraints: two on the norm of the drivings and one on the entropy production rate.}. We exactly enforce the two normalization conditions on the $g_n^{(l)}$ coefficients expressing them as a function of unconstrained real parameters $\{\tilde{g}_n^{(l)} \}$ through the relations
\be 
	g_n^{(l)} =  |g^{(l)}|  \frac{\tilde{g}_n^{(l)}}{ \sqrt{ \sum_m \left[\tilde{g}_m^{(l)}\right]^2 } },
	\label{eq:gn_tilde}
\ee
and then we perform the numerical optimization with respect to $\{\tilde{g}_n^{(l)} \}$. Within the Section, the coefficients $g_n^{(l)}$ are always to be intended as functions of $\{ \tilde{g}_n^{(l)} \}$ through Eq.~(\ref{eq:gn_tilde}).

The constraint on the entropy production rate is instead imposed through a Lagrange multiplier. We thus introduce the Lagrangian
\be
	 L(\{ \tilde{g}_n^{(l)} \}, \alpha) = P(\{g_n^{(l)}\}) +\alpha\left[\sigma(\{{g}_n^{(l)}\}) - \sigma_i\right],
\ee
where $\alpha$ is a Lagrange multiplier. Here the power $P(\{{g}_n^{(l)}\})$ and entropy production  $\sigma(\{{g}_n^{(l)}\})$ are computed as in Eq.~(\ref{eq:p_sigma_def}) considering only real coefficients (as proven in the previous Section, this is not a restrictive assumption). As shown in Ref.~\cite{platt1987}, we can numerically solve the optimization problem performing a gradient descent of $L(\{ \tilde{g}_n^{(l)} \}, \alpha)$ with respect to the $\tilde{g}_n^{(l)}$ parameters, and a gradient ascent with respect to $\alpha$. However, this can lead to an unstable optimization. Ref.~\cite{platt1987} proposes to overcome this problem adding an additional penalty term to the Langrangian, leading to
\begin{multline}
	 L(\{ \tilde{g}_n^{(l)} \}, \alpha) = P(\{g_n^{(l)}\}) +\alpha\left[\sigma(\{{g}_n^{(l)}\}) - \sigma_i\right] 
	 \\+ \gamma \left[    \sigma(\{{g}_n^{(l)}\}) - \sigma_i  \right]^2,
	 \label{eq:app_lagrange_penalty}
\end{multline}
where $\gamma>0$ is a fixed hyperparameter. This additional term has no effect when the constraint is exactly satisfied, but improves convergence (see Ref.~\cite{platt1987} for details).

In practice, we use the PyTorch framework~\cite{paszke2019} to compute the gradients of $L(\{ \tilde{g}_n^{(l)} \}, \alpha)$, given in Eq.~(\ref{eq:app_lagrange_penalty}), with respect to $\{ \tilde{g}_n^{(l)} \}$ and $\alpha$ using automatic differentiation. We then use the Adam optimizer~\cite{kingma2014}, with a learning rate of $0.01$, to minimize $L(\{ \tilde{g}_n^{(l)} \}, \alpha)$ with respect to $\{ \tilde{g}_n^{(l)} \}$, and we use a gradient ascent, with a learning rate of $0.003$, to maximize $L(\{ \tilde{g}_n^{(l)} \}, \alpha)$ with respect to $\alpha$. To ensure that $\alpha$ remains positive~\cite{platt1987}, we express it as $\alpha = e^\beta$, and perform a gradient ascent with respect to $\beta$. We also fix $\gamma=1$.

We start from a random guess of $\{ \tilde{g}_n^{(l)} \}$ uniformly sampled in the $[0,1)$ interval, and we start from $\alpha=1$. We then alternate one optimization step for $\{ \tilde{g}_n^{(l)} \}$, and one for $\alpha$, for a total of $16.000$ times. We verified that this choice of hyperparameters and this number of optimization steps lead to a very stable and consistent optimization. The same hyperparameters are used for all results presented in this manuscript.

We now show some representative values of $g_n^{(l)}$ that emerge from the numerical optimizations leading to Figures~\red{4} and \red{5} (main text). 
\begin{figure}[ht] 
    	\centering
    	\includegraphics[width=1.\linewidth]{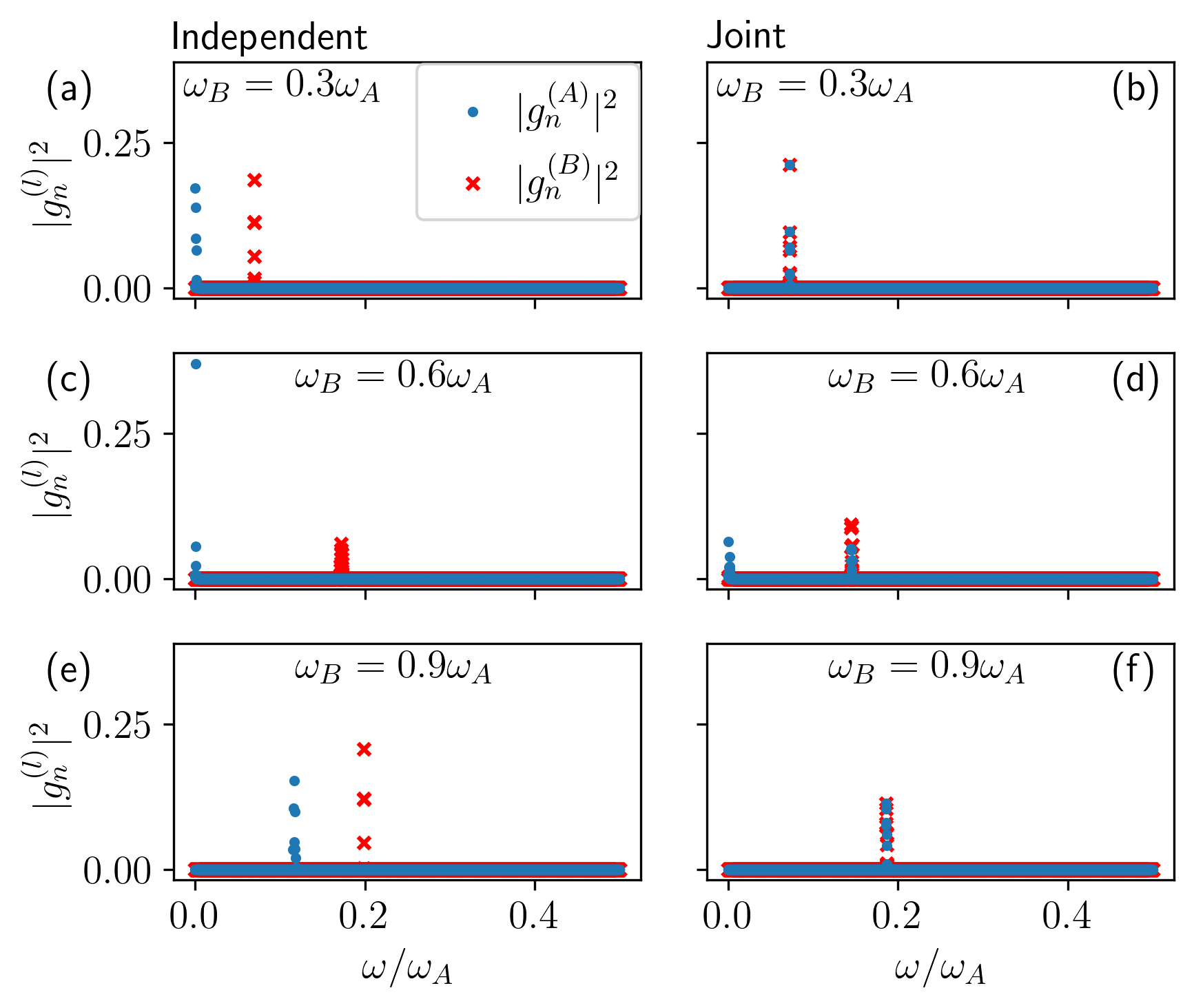}
    	\caption{$|g_n^{(l)}|^2$, for $l=A$ (blue circles) and $l=B$ (red crosses), as a function of the frequency $\omega=n\Omega$, found in specific points of Figure~\red{4} (main text), \ie in the moderate damping case. The left (right) column is relative to the independent (joint) case, while each row corresponds to a different value of $\omega_B$ reported on each panel. (a,b,e,f) correspond to the maximum power driving, while (c) corresponds to the $(\tilde{\sigma},|\tilde{P}|) = (0.13,0.021)$ point, and (d) to $(\tilde{\sigma},-\tilde{P})=(0.12,0.065)$.}
\label{fig:histo_01}
\end{figure}
In Supplementary Figure~\ref{fig:histo_01} we report the square modulus of the optimal driving coefficients $|g_n^{(l)}|^2$, as a function of the frequency $\omega=n\Omega$, relative to specific points in Figure~\red{4} (main text). In particular, the left (right) column is relative to the independent (joint) case, while each row corresponds to a different value of $\omega_B$ (reported on each panel). Panels (a,b,e,f) report the square modulus of the coefficients in the maximum power case, while panel (c,d) report a case with low power and low entropy production. In particular, Panel (c) corresponds to the $(\tilde{\sigma},|\tilde{P}|) = (0.13,0.021)$ point, while Panel (d) corresponds to $(\tilde{\sigma},|\tilde{P}|)=(0.12,0.065)$. Since we are optimizing over 5.000 values of $\omega$, points that appear to be ``vertical'' correspond to extremely similar frequencies, and thus we consider them as a single frequency (it is likely that optimizing for more steps or decreasing the learning rate could yield exactly a single frequency). As discussed in the main text, we see that the independent case always finds two distinct frequencies for each QHO [see panels (a,c,e)]. On the other hand, the joint case at maximum power is always given by a single frequency [see panels (b,f)]. Along the rest of the Pareto front, the joint case is also always optimal at a single frequency, except for the low power regime at $\omega_B=0.6\omega_A$, where two frequencies become optimal. Such an example is reported in Panel (d). Interestingly, as opposed to the independent case where each QHO is operated at a distinct frequency, here QHO $A$ is driven at both frequencies.

\begin{figure}[ht] 
    	\centering
    	\includegraphics[width=1.\linewidth]{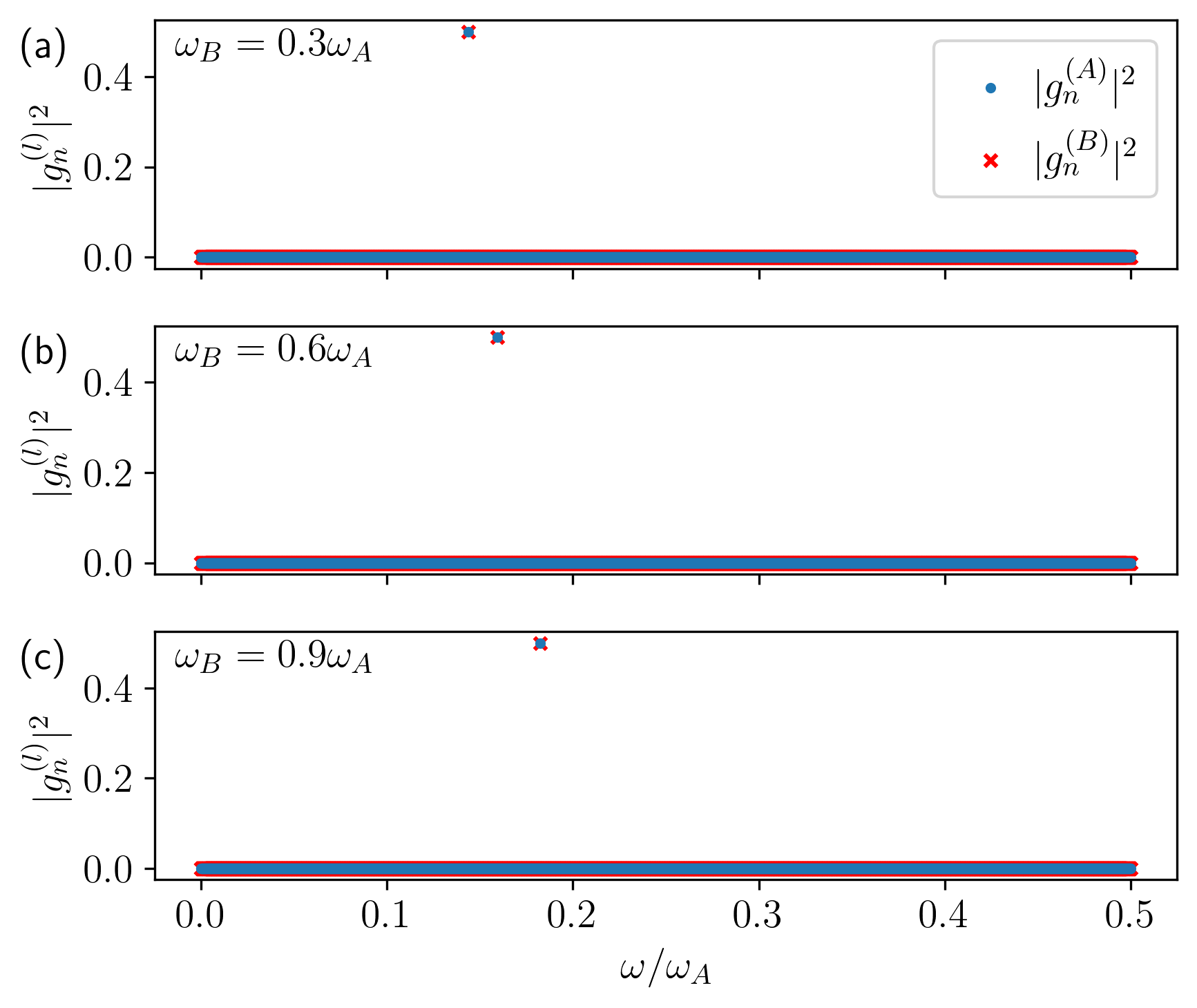}
    	\caption{$|g_n^{(l)}|^2$, for $l=A$ (blue circles) and $l=B$ (red crosses), as a function of the frequency $\omega=n\Omega$, found in specific points of Figure~\red{5} (main text), \ie in the ultra-strong damping case. Only the joint case is considered since the independent case does not operate as a heat engine in this regime. Each row corresponds to a different value of $\omega_B$ reported on each panel. All plots correspond to the driving that maximizes the power.  }
\label{fig:histo_100}
\end{figure}
In Supplementary Figure~\ref{fig:histo_100} we report the same data, \ie the square of the driving coefficients as a function of the frequency, relative to the maximum power points in Figure~\red{5} (main text). Only the joint case is analyzed since the independent case does not operate as a heat engine in this regime. Each row corresponds to a different value of $\omega_B$ (reported on each panel). As discussed in the main text and confirmed in Figure~\red{5} (main text), in all these cases the optimal driving is monochromatic and occurs at $\omega=\bar{\omega}$. 
\section*{Supplementary Note 6: Logarithmic negativity}
\label{app:en}
In this Section we provide some details on the evaluation of possible fingerprint of entanglement in the WM in the case of bath-mediated correlations. To this end we restrict the analysis to the $\nu=2$ reservoir, the one responsible for mediating correlations, since the other is only weakly coupled to the quantum system.

We recall that the logarithmic negativity~\cite{simon2000, serafini2004, paz09, correa12} is defined as
\be
\label{eq:en_def}
E_n \equiv {\rm Max} \big[0, - \log(2\tilde{\nu})\big]~.
\ee
A finite $E_n$ is a fingerprint of entanglement. Therefore it is required that $\tilde{\nu} < 1/2$, where $\tilde{\nu}$ is the so-called symplectic eigenvalue of the partial transposed density matrix~\cite{simon2000}.
For gaussian states this quantity can be written as~\cite{paz09, correa12}
\be
\label{eq:nutilde_def}
2\tilde{\nu}^2= \Delta -\sqrt{\Delta^2 - 4 \det(\Sigma)}~.
\ee
In the above equation we have introduced the covariance matrix
\beq 
\Sigma= \left(\begin{array}{cc}
\alpha & \gamma \\
\gamma^t & \beta
\end{array}\right)
\eeq
where the two-by-two submatrices $\alpha $ and $\beta$ refer to the $A$ and $B$ QHO correlators, while $\gamma$ refers to mixed ones.
In particular, these are given by
\beq
\alpha= \left(\begin{array}{cc}
\langle x_A^{(0)}(t)x_A^{(0)}(t)\rangle_s & \langle x_A^{(0)}(t)p_A^{(0)}(t)\rangle_s \\
 \langle p_A^{(0)}(t)x_A^{(0)}(t)\rangle_s & \langle p_A^{(0)}(t)p_A^{(0)}(t)\rangle_s 
 \end{array}\right)
\eeq
\beq
\beta = \left(\begin{array}{cc}
\langle x_B^{(0)}(t)x_B^{(0)}(t)\rangle_s & \langle x_B^{(0)}(t)p_B^{(0)}(t)\rangle_s \\
 \langle p_B^{(0)}(t)x_B^{(0)}(t)\rangle_s & \langle p_B^{(0)}(t)p_B^{(0)}(t)\rangle_s 
 \end{array}\right)
\eeq
and
\beq
\gamma = \left(\begin{array}{cc}
\langle x_A^{(0)}(t)x_B^{(0)}(t)\rangle_s & \langle x_A^{(0)}(t)p_B^{(0)}(t)\rangle_s \\
 \langle p_A^{(0)}(t)x_B^{(0)}(t)\rangle_s & \langle p_A^{(0)}(t)p_B^{(0)}(t)\rangle_s 
 \end{array}\right)
\eeq
in terms of equal time symmetrized (see the subscript\red{$s$}) position and momentum correlators. Here, we are interested in the long time behaviour, when a stationary regime has been reached, therefore these quantities do not depend on time.
In Eq.~\eqref{eq:nutilde_def} we have also introduced
\be
\Delta = \det(\alpha) + \det(\beta) - 2 \det(\gamma).
\ee
The symmetrized position and momentum correlators can be written in terms of the imaginary part of the response function ${\bm \chi_2}$ as
\begin{align} 
\label{eq:xx_pp_correlators} 
&\langle x^{(0)}_l (t) x^{(0)}_{l'}(t)\rangle_s = \int_{-\infty}^{+\infty}\frac{\d\omega}{2\pi m} {\chi_2^{(ll')}}''(\omega) \coth\left(\frac{\omega}{2T_2}\right)\nonumber \\
&\langle p^{(0)}_l (t) p^{(0)}_{l'}(t)\rangle_s = m\int_{-\infty}^{+\infty}\frac{\d\omega}{2\pi } {\omega^2 \chi_2^{(ll')}}''(\omega) \coth\left(\frac{\omega}{2T_2}\right),
\end{align}
while all other correlators of the form $\langle x^{(0)}_l (t) p^{(0)}_{l'}(t)\rangle_s$ are zero. Therefore, all submatrices $\alpha$, $\beta$, and $\gamma$ have diagonal components only. It is worth to note that the momentum-momentum correlator in Eq.~\eqref{eq:xx_pp_correlators} contains a logarithmic divergence at high frequencies and, in general, a cut-off $\omega_c$ in the Ohmic Drude form of the damping function $\gamma_2(\omega)$. However, it is possible to demonstrate (not shown) that $\tilde{\nu}$, and thus also $E_n$, have finite values in the $\omega_c\to +\infty$ limit. Indeed, even though $\tilde{\nu}$ depends on combinations of position and momentum correlators, all logarithmic divergences cancel out and thus $\tilde{\nu}$ is a well-behaved expression also when $\omega_c\to + \infty$.

\begin{figure}[!t] 
    	\centering
    	\includegraphics[width=1.0\linewidth]{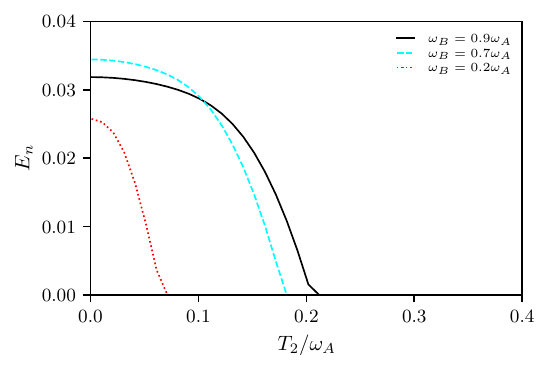}        	
  \caption{ Logarithmic negativity $E_n$ as a function of the normalized temperature $T_2/\omega_A$ for $\gamma_2=0.1\omega_A$. Different curves refer to different values of $\omega_B/\omega_A$.
\label{fig:en}}
\end{figure}

Inspecting Eqs.\eqref{eq:nutilde_def}-\eqref{eq:xx_pp_correlators}, one can evaluate $E_n$ in Eq.~\eqref{eq:en_def}. A representative example of $E_n$ for the moderate damping case $\gamma_2=0.1\omega_A$, and for different $\omega_B/\omega_A$, is reported in Supplementary Figure~\ref{fig:en} showing that it is finite at sufficiently low temperatures, while it vanishes at higher ones.
A closed expression for $\tilde{\nu}$ can be found in the strong damping regime. 
Indeed, exploiting the asymptotic expressions reported \red{in the main text} one can arrive at analytical expressions for the correlators in Eq.~\eqref{eq:xx_pp_correlators}. In doing so, a useful relation is
\be
\lim_{\eta_0\to 0} \omega \coth\left(\frac{\omega}{2T_2}\right)\frac{\eta_0}{\omega^2+\eta_0^2}\to 2\pi T_2 \delta(\omega)~,
\ee
 which implies that the position-position correlator takes contributions both from $\omega\sim 0$ and from $\omega \sim \bar{\omega}$, while the momentum-momentum one is governed by contributions close to $\bar{\omega}$. After some passages, one eventually arrives at
\be
\tilde{\nu}^2=\frac{T_2 \bar{\omega}^3\coth^2(\bar{\omega}/(2T_2))}{2 \omega_A^2\omega_B^2}\frac{1}{\coth(\bar{\omega}/(2T_2))+\frac{T_2 (\omega_B^2-\omega_A^2)^2}{2 \bar{\omega} \omega_A^2\omega_B^2}}\, .
\label{app:nutilde2_final}
\ee
Recalling that finite entanglement is expected for $\tilde{\nu} <1/2$, this will define the critical temperature $T^*$ in the asymptotic regime of large $\gamma_2$: $\tilde{\nu}(T^*)=1/2$.